\pdfoutput=1

\documentclass[fleqn,usenatbib,usedcolumn]{mnras}
\usepackage{natbib}

   \usepackage[british]{babel}             
   \usepackage{newtxtext}                  
   \usepackage[slantedGreek]{newtxmath}    
  
   \let\la=\lesssim 
  
   \usepackage[T1]{fontenc}               
   \usepackage{graphicx}

\def\gtsim {>\kern-1.2em\lower1.1ex\hbox{$\sim$}~}   % Greater than sim
\def\ltsim {<\kern-1.2em\lower1.1ex\hbox{$\sim$}~}   % Less than sim

\usepackage{graphicx}	% Including figure files
\usepackage{amsmath}	% Advanced maths commands
\usepackage{amssymb}	% Extra maths symbols

\title[Stellar migrations and metal flows in a MW-type galaxy]{
Stellar migrations and metal flows -- Chemical evolution of the thin disc of a simulated Milky Way analogous galaxy}
\author[F. Vincenzo \& C. Kobayashi]{Fiorenzo Vincenzo$^{1,2}$\thanks{email: vincenzo.3@osu.edu} \& 
Chiaki Kobayashi$^{3}$\thanks{email: c.kobayashi@herts.ac.uk}
\\
$^{1}$Center for Cosmology and AstroParticle Physics, The Ohio State University, 191 West Woodruff Avenue, Columbus, OH 43210, USA \\ 
$^{2}$Department of Astronomy, The Ohio State University, 140 West 18th Avenue, Columbus, OH 43210, USA \\ 
$^{3}$Centre for Astrophysics Research, University of Hertfordshire, College Lane, Hatfield, AL10 9AB, UK  }

\begin{document}

\date{Accepted 2020 May 21. Received 2020 May 21; in original form 2020 February 21}

\pagerange{\pageref{firstpage}--\pageref{lastpage}} \pubyear{2020}

\maketitle

\label{firstpage}

%%%%%%%%%%%%%%%%%%%%%%%%%%%%%%%%%%%%%%
%%%%%%%%%%%%%%%%%%%%%%%%%%%%%%%%%%%%%%

\begin{abstract}
In order to understand the roles of metal flows in galaxy formation and evolution, we analyse our self-consistent cosmological chemo-dynamical simulation of a Milky Way like galaxy during its thin-disc phase. Our simulated galaxy disc qualitatively reproduces the variation of the dichotomy in [$\alpha$/Fe]--[Fe/H] at different Galactocentric distances as derived by APOGEE-DR16, as well as the stellar age distribution in [$\alpha$/Fe]--[Fe/H] from APOKASC-2. 
The disc grows from the inside out, with a radial gradient in the star-formation rate during the entire phase. Despite the radial dependence, the outflow-to-infall ratio of metals in our simulated halo shows a time-independent profile scaling with the disc growth.
The simulated disc undergoes two modes of gas inflow:
\textit{(i)} an infall of metal-poor and relatively low-[$\alpha$/Fe] gas,
and \textit{(ii)} a radial flow where already chemically-enriched gas moves inwards with an average velocity of $\sim0.7$ km/s. Moreover, we find that stellar migrations mostly happen outwards, on typical time scales of $\sim5$ Gyr. Our predicted radial metallicity gradients agree with the observations from APOGEE-DR16, and the main effect of stellar migrations is to flatten the radial metallicity profiles by 0.05 dex/kpc in the slopes. 
We also show that the effect of migrations can appear more important in [$\alpha$/Fe] than in the [Fe/H]--age relation of thin-disc stars.
\end{abstract}

%%%%%%%%%%%%%%%%%%%%%%%%%%%%%%%%%%%%%%
%%%%%%%%%%%%%%%%%%%%%%%%%%%%%%%%%%%%%%


 %%%%%%%%%%%%%%%%%%%%%%%%%%%%%%%%%%%%%%
 %%%%%%%%%%%%%%%%%%%%%%%%%%%%%%%%%%%%%%

\section{Introduction} \label{sec:intro}

Chemical abundances in the stars and interstellar medium (ISM) can be used to understand the formation and evolutionary histories of galaxies (see, for example, \citealt{tinsley1980,matteucci2001,matteucci2012,kobayashi2006,pagel2009}). The age--metallicity relation which is observed among the stars in the Milky Way (MW) contains information about \textit{(i)} the efficiency of star formation and accretion timescale of gas from the environment at different Galactocentric distances \citep{chiappini1997,boissier2000,prantzos2000,molla2016}; \textit{(ii)} radial migration of stars from one Galaxy region to another \citep{schoenrich2009,spitoni2015,buck2020}; \textit{(iii)} the delay-time distribution function of Type Ia Supernovae (SNe; the main producers of iron-peak elements in the cosmos) \citep{matteucci1986}; and \textit{(iv)} the relative role of low-mass and massive stars in the initial mass function (IMF) \citep{romano2005}. All these aspects have been extensively investigated in the past, by means of analytical and semi-analytical chemical-evolution models.

The [$\alpha$/Fe]--[Fe/H] diagram is an important chemical abundance diagnostic not only for our Galaxy but also for external galaxies \citep{kobayashi2016}. The [$\alpha$/Fe] ratios depend on the star formation history (SFH; \citealt{matteuccibrocato1990}), IMF \citep{romano2005}, and on the progenitor models of Type Ia SNe \citep{kobayashi1998,matteuccirecchi2001,kobayashi2009}, as well as on the nucleosynthesis yields from massive stars, which explode as core-collapse SNe \citep{kobayashi2006,romano2010,prantzos2018}. 

In our Galaxy, the time evolution of the radial chemical abundance gradients can be obtained by using different stellar tracers on the Galaxy disc \citep{anders2017b}, and it can depend on the effects of radial migrations \citep{schoenrich2009,minchev2018}, infall of gas \citep{hou2000,cescutti2007,schoenrich2017}, and radial gas flows \citep{laceyfall1985,portinari2000,spitoni2011,bilitewski2012,pezzulli2016}, as shown by many chemical-evolution models and chemical abundance measurements over the years \citep[e.g.,][]{molla1997,magrini2009,magrini2016,magrini2017,stanghellini2010,stanghellini2018,molla2015,molla2016,molla2019,molla2019b}.

\begin{figure}
\centering
\includegraphics[width=8.0cm]{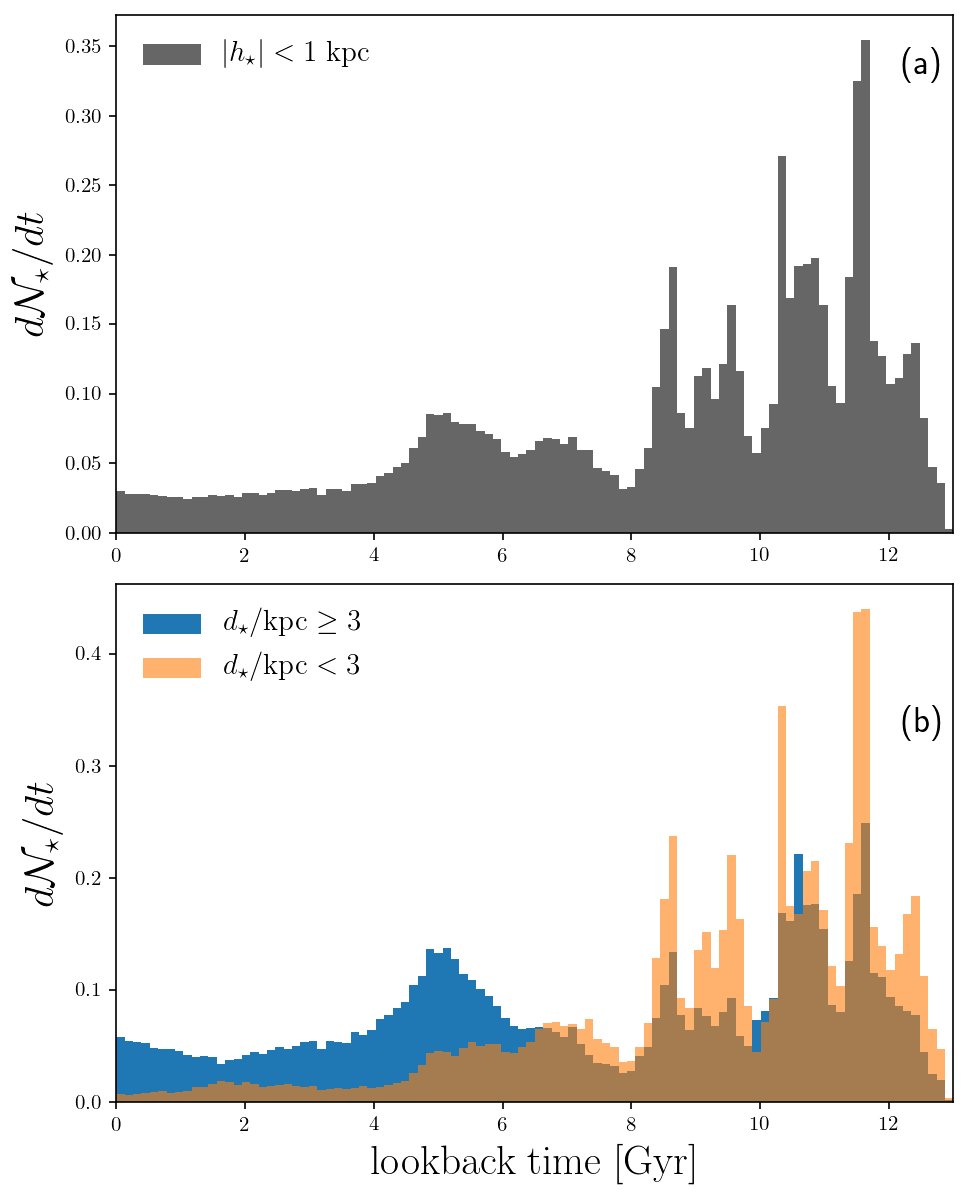} 
\caption{ \textbf{\textit{(a)}} The age distribution of the stars at the present-time in the simulated galaxy disc. \textbf{\textit{(b)}} Orange area: the age distribution of the stars at the present-time in the galaxy disc with galactocentric distance $d_{\star} < 3\;\text{kpc}$; blue area: the age distribution of the stars at the present-time in the galaxy disc with galactocentric distance $d_{\star} \ge 3\;\text{kpc}$. }
\label{fig1}
\end{figure}

\begin{figure}
\centering
\includegraphics[width=8.0cm]{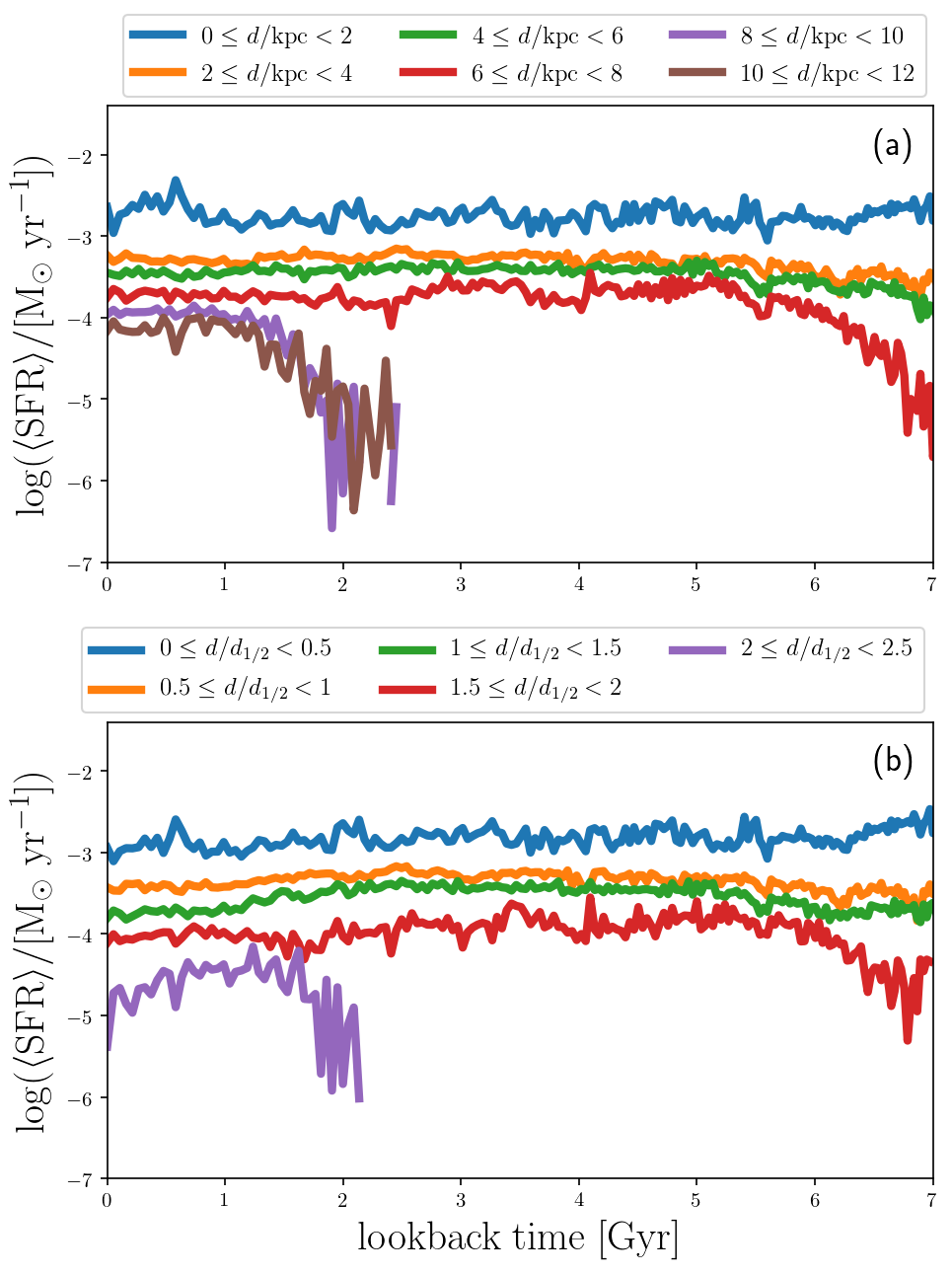} 
\caption{\textbf{\textit{(a)}} The propagation of the star formation activity on the simulated galaxy disc as a function of the look-back time, by considering all available snapshots; the various curves with different colours correspond to different galactocentric annuli, $d$, in physical units (kpc); \textbf{\textit{(b)}} the evolution of the SFR in the galaxy disc as a function of the look-back time, by dividing the disc in many concentric radial intervals normalised by the half-mass radius, $d_{1/2}$, at each redshift. }
\label{fig2}
\end{figure}

Many recent MW observational surveys have been able to integrate the chemical abundance information for a large number of MW stars with other important observables, like position and kinematics from Gaia (e.g., \citealt{sanders2018,grand2018b,schoenrich2019,leung2019b,khoperskov2019}), and 
asteroseismic measurements of the internal structural properties of the stars from space missions like Kepler, K2, and TESS (e.g., \citealt{miglio2013,casagrande2016,anders2017a,silvaaguirre2018,pinsonneault2018,rendle2019,chaplin2020}). Finally, spectroscopic surveys like Gaia-ESO  (e.g., \citealt{hayden2018,magrini2018}), GALAH (e.g., \citealt{buder2018,hayden2019,griffith2019,lin2019}), APOGEE (e.g., \citealt{hayden2014,weinberg2019,leung2019a}), and LAMOST (e.g., \citealt{xiang2019}) have been able to provide very large statistical samples of stars with precise chemical abundance measurements to study the formation and evolution of our Galaxy like never before. 

Examples of physical and dynamical processes recently explored to explain some MW survey observational data are, for example, the effect of cosmological mergers and satellite impacts (see, for example, the observational works of \citealt{belokurov2018,helmi2018,gallart2019,belokurov2019,chaplin2020}, as well as \citealt{laporte2019a,laporte2019c,bignone2019,brook2020,grand2020} for theoretical works); disc heating and perturbation mechanisms (e.g., \citealt{antoja2018, mackereth2019, dimatteo2019, laporte2019b, fragkoudi2019, belokurov2019,amarante2020}); stellar migrations (e.g., \citealt{kubryk2015a,kubryk2015b,minchev2018,frankel2018,feltzing2019}); gas and fountain flows \citep{spitoni2019,grand2018a,grand2019}; and intrinsic inhomogeneous chemical enrichment \citep{kobayashi2011}. 

Concerning the chemical abundance data, from the point of view of chemical-evolution models, it is not clear yet which of these processes is dominant in determining the observed radial metallicity gradient in the stellar disc of our Galaxy; moreover, it is of paramount importance to determine the impact of stellar migrations and gas infall on the observed [$\alpha$/Fe]--[Fe/H] and age--metallicity relation among the thin-disc stars of our Galaxy.
Chemodynamical simulations embedded in a cosmological framework can, in principle, provide a self-consistent chemical-evolution model including {\it all} of the these physical and dynamical processes that determine the evolution of chemical abundances in galaxies: stellar nucleosynthesis by different kinds of stars and SNe in the cosmos, gas flows, stellar migrations, and cosmological growth 
(e.g., \citealt{kobayashi2011,few2014,maio2015,ma2017a,ma2017b,vincenzo2018b,grand2018a,clarke2019,torrey2019,valentini2019,buck2020}). 
For a different approach, in which both cosmological growth and chemical evolution are taken into account but not (hydro)dynamical processes, the work of \citet{calura2009} represents the first chemical-evolution model, embedded in a hierarchical semi-analytical model, that predicted the currently observed bimodal distribution in [$\alpha$/Fe]--[Fe/H] between thick- and thin-disc stars in our Galaxy \citep{hayden2014,weinberg2019}, even though galaxies are not spatially-resolved in \citet{calura2009}.

The main scope of this paper is to use our cosmological zoom-in  chemodynamical simulation of a MW-type galaxy to understand the role of stellar migrations and gas flows in the chemical evolution of the thin-disc of the MW. This paper is organised as follows;  Section \ref{sec:model} gives brief description of our simulation; in Section \ref{sec:results} we presents the results of our analysis; finally, in Section \ref{sec:conclusions} we draw our conclusions.

%%%%%%%%%%%%%%%%%%%%%%%%%%%%%%%%%%%%%%%%%%%%%%%%%%%%%%%%%
%%%%%%%%%%%%%%%%%%%%%%%%%%%%%%%%%%%%%%%%%%%%%%%%%%%%%%%%%
%%%%%%%%%%%%%%%%%%%%%%%%%%%%%%%%%%%%%%%%%%%%%%%%%%%%%%%%%
%%%%%%%%%%%%%%%%%%%%%%%%%%%%%%%%%%%%%%%%%%%%%%%%%%%%%%%%%
%%%%%%%%%%%%%%%%%%%%%%%%%%%%%%%%%%%%%%%%%%%%%%%%%%%%%%%%%
%%%%%%%%%%%%%%%%%%%%%%%%%%%%%%%%%%%%%%%%%%%%%%%%%%%%%%%%%
%%%%%%%%%%%%%%%%%%%%%%%%%%%%%%%%%%%%%%%%%%%%%%%%%%%%%%%%%
%%%%%%%%%%%%%%%%%%%%%%%%%%%%%%%%%%%%%%%%%%%%%%%%%%%%%%%%%

\begin{figure*}
\centering
\includegraphics[width=16.4cm]{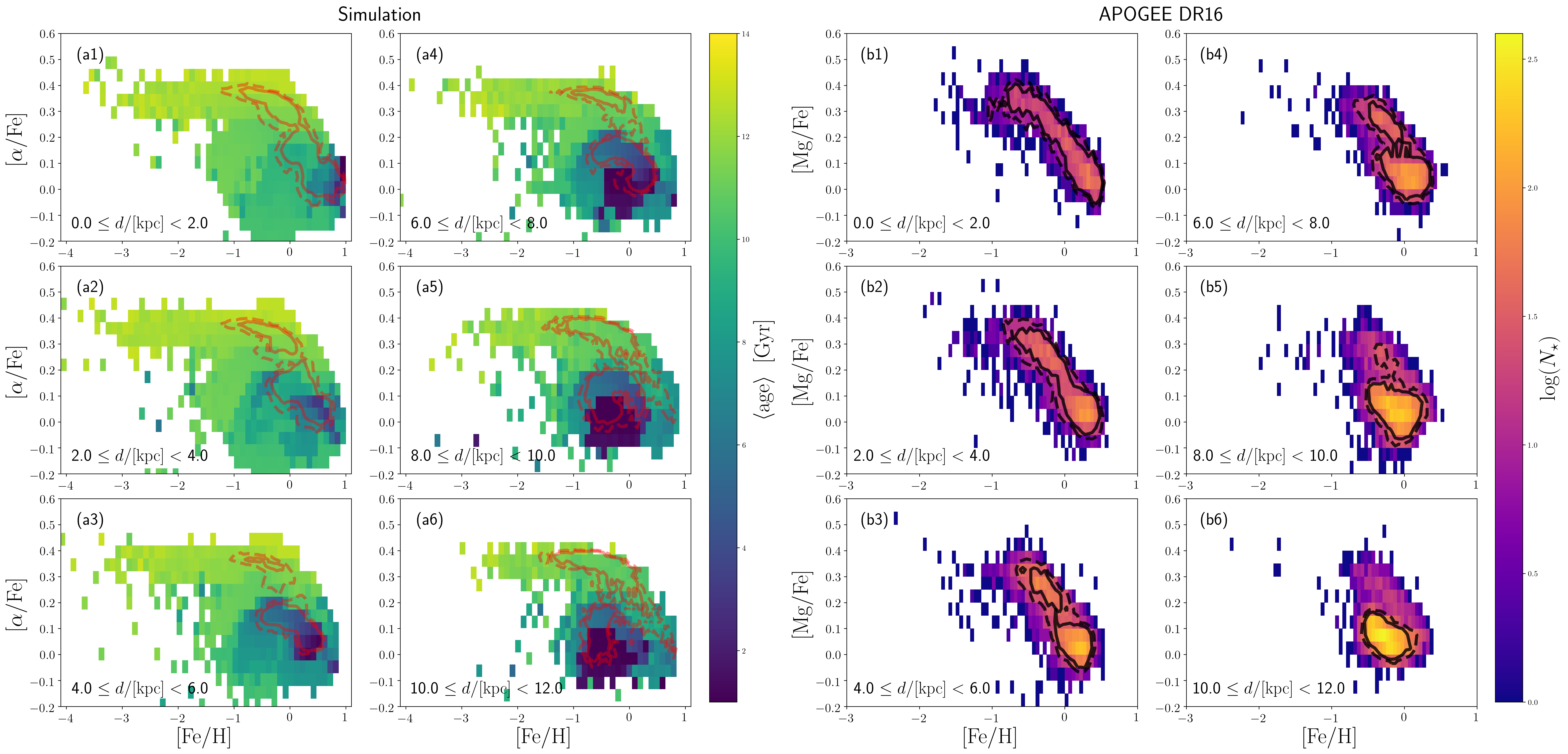} 
\caption{ \textbf{\textit{(Left panels)}}: the predicted [$\alpha$/Fe] versus [Fe/H] diagram by considering all the stellar populations in the simulated galaxy disc at the present-time. Different panels correspond to different bins of present-day galactocentric distances, and the colour-coding represents the average stellar age, whereas the red contours show the 10 and 20 per cent levels in the number of stars as normalised with respect to the maximum of the 2-D distribution; \textbf{\textit{(Right panels)}}: the observed [Mg/Fe] versus [Fe/H] from APOGEE-DR16 \citep{ahumada2019} in different bins of Galactocentric distances, $d$; the colour-coding corresponds to the logarithm of total  number of stars within each bin of the 2-D distribution. }
\label{fig:alphafe-simulation}
\end{figure*}

\begin{figure}
\centering
\includegraphics[width=8.0cm]{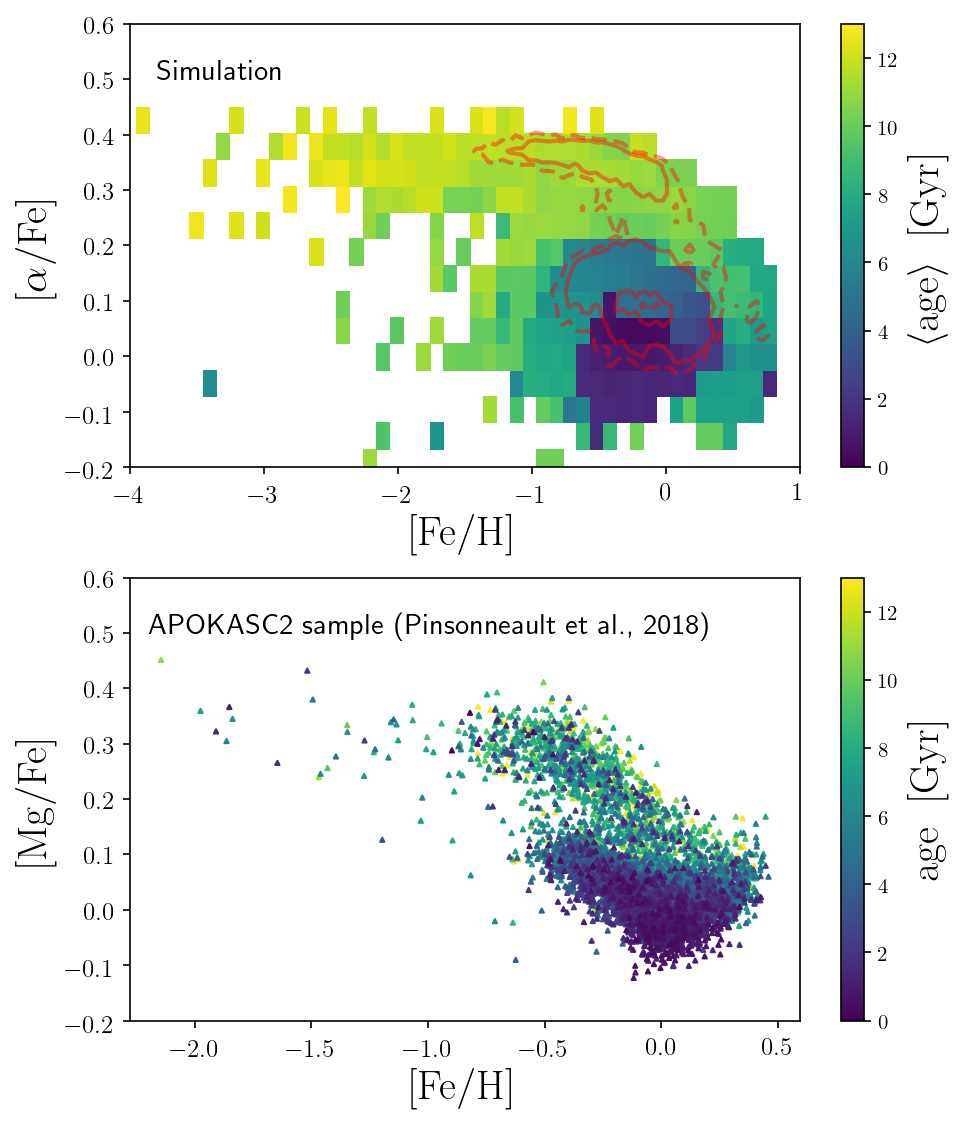} 
\caption{ \textit{\textbf{(Top panel)}} The predicted age distribution in [$\alpha$/Fe]--[Fe/H] of all stars in the simulated galaxy disc at the present. The colour coding represents the average age in each bin of the 2D histogram, whereas the solid and dashed contours in red highlight the 10 and 20 per cent levels, respectively, in the number of stars populating the diagram. \textit{\textbf{(Bottom panel)}} The observed age distribution of the stars from the second APOKASC catalog \citep{pinsonneault2018} in the [$\alpha$/Fe]--[Fe/H] diagram (chemical abundances from APOGEE-DR16).  }
\label{fig:apokasc}
\end{figure}

\begin{figure*}
\centering
\includegraphics[width=16.4cm]{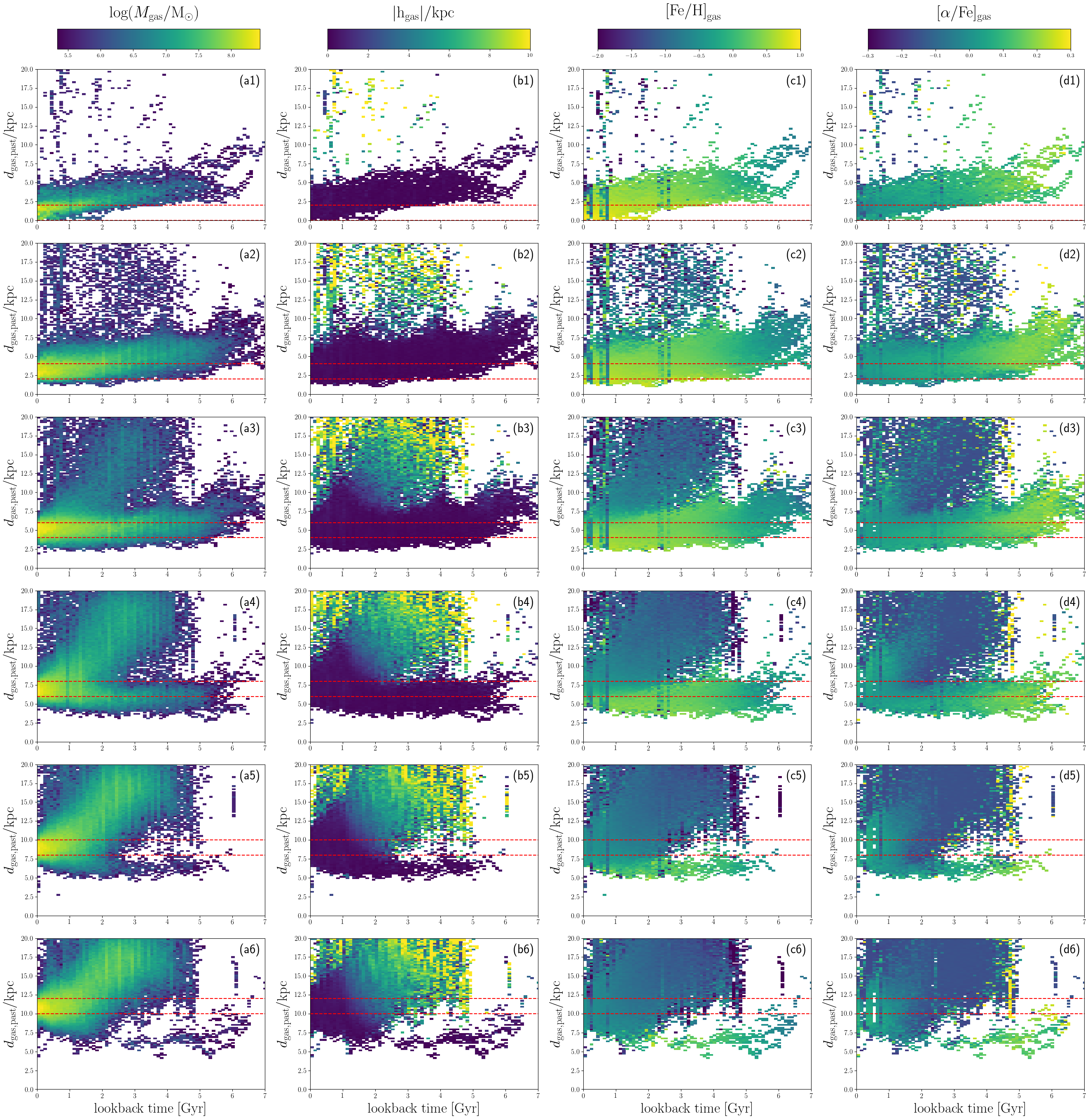} 
\caption{Gas infall and radial flow history that determines the formation and evolution of the present-day simulated gaseous disc. All panels show how the galactocentric distance, $d_{\text{gas,past}}$, of all the gas particles that today reside on the disc evolves as a function of look-back time. Within each subgroup of figures (different columns), each row corresponds to a different present-day galactic annulus with length $\Delta R = 2\;\text{kpc}$, as marked by the horizontal red dashed lines. The different columns show different average quantities computed in the $d_{\text{gas,past}}$ versus \textit{look-back time} diagram: average gas mass in each 2-D grid element (a1-a6); galactic height, below or above the disc, $|h_{\rm gas}|$ (b1-b6); [Fe/H] (c1-c6); and  [$\alpha$/Fe] (d1-d6). }
\label{fig:gas-flow_thin-disc}
\end{figure*}

\section{The simulation} \label{sec:model}

The simulated galaxy is run with our chemodynamical code \citep{kobayashi2007,vincenzo2018a,vincenzo2018b,vincenzo2019}, based on Gadget-3 \citep{springel2005}, but including all relevant baryon physics: radiative cooling depending on metallicity and [$\alpha$/Fe], star formation, thermal feedback from stellar winds and SNe, and chemical enrichment of all elements up to Zn from asymptotic giant-branch (AGB) stars, Type Ia SNe, and core-collapse SNe. Up to 50 per cent of stars with initial masses $\ge 20\;M_{\sun}$ is assumed to be hypernovae (depending on metallicity; see \citealt{kobayashi2011} for more details), which produce more energy ($>10^{51}$ erg) and iron. In the following subsections we summarize the main assumptions and characteristics of our simulation that are relevant to this paper; for more details about the simulation code, we address the readers to the works of \citet{kobayashi2004,kobayashi2007,kobayashi2011,kobayashi2014,taylor2014,vincenzo2018a,vincenzo2018b,vincenzo2019}, and \citet{haynes2019}. 

\subsection{Initial conditions}
The initial conditions of our simulations are taken from the Aquarius Project \citep{springel2008}, labeled as \textit{Aq-C-5}, which give rise to a MW-sized DM halo at redshift $z=0$ \citep{scannapieco2012}. We assume the $\Lambda$-cold dark-matter Universe with the following cosmological parameters: $\Omega_{0}=0.25$, $\Omega_{\Lambda}=0.75$, $\Omega_{b}=0.04$, $H_{0}=100\times h = 73\;\text{km}\,\text{s}^{-1}\,\text{Mpc}^{-1}$, which are 
consistent with the one- and five-year Wilkinson Microwave Anisotropy Probe \citep{komatsu2009}. 
The resolution in mass of our simulation is $M_{\text{DM}} \approx 1.576 \times 10^{6}\;h^{-1}\,\text{M}_{\sun}$ for dark-matter (DM) particles, and $M_{\text{gas}} \approx 3.0 \times 10^{5}\;h^{-1}\,\text{M}_{\sun}$ for gas particles in the initial condition, with a total number of DM and gas particles which is $N_{\text{DM}} = 1,612,268$ and $N_{\text{gas}} = 1,623,903$, respectively. Finally, the gravitational smoothing length is $\epsilon_{\text{gas}}=0.5\;h^{-1}\,\text{kpc}$ in comoving units. 

 We choose this galaxy, because the initial conditions give rise to a marked dichotomy between thick- and thin-disc stars in the chemical abundance space at the present-time, in agreement with APOGEE observations \citep{weinberg2019}, and because the simulated galaxy undergoes a significant merger event at redshift $z\sim3$, which almost completely destabilises the gaseous disc, which was present in the galaxy since $z\sim4$ (see Kobayashi 2020, in prep. for the other galaxies). Many observational works have recently found evidence of a merger event that our Galaxy suffered at high redshift ($z \sim 2$; see, for example, \citealt{chaplin2020}, even though it is not easy to date directly any high-redshift merger by looking at the present-day substructures in the MW halo) with a companion galaxy which is now called Gaia-Sausage or Gaia-Enceladus (e.g., \citealt{belokurov2018,helmi2018,gallart2019,chaplin2020}, but see also \citealt{iorio2019,vincenzo2019a,mackereth2020}).
After the merger, the simulated galaxy develops a new gaseous disc, which grows in mass and size from $z\approx 2.6$ down to $z=0$. In this work we focus our analysis on the chemical evolution of our simulated galaxy after the merger event. 

These initial conditions were selected for the Aquila code comparison project in \citet{scannapieco2012} because it has a relatively quiet formation history, and is mildly isolated at $z = 0$, with no massive neighbouring halo.
These initial conditions have been used in many chemodynamical simulations; in the Aquila code comparison, our code is labelled as G3-CK. The simulation presented in the Aquila project was with Salpeter IMF, and the simulation presented in \citet{haynes2019} includes neutron-capture processes.
However, the basic properties of the simulated galaxy of this paper are very similar to those of the G3-CK simulation.
Compared to other simulated galaxies, the G3-CK galaxy lies in the middle of the scattered distributions for the morphology, stellar circularity, circular velocity curve, stellar mass, gas fraction, size, and star formation history.
Although the stellar mass was still larger and the size was smaller than observed at a give halo mass, our thermal SN feedback scheme is acceptable because of the inclusion of hypernovae which, on average, deposit three times more energy in the surrounding gas particles than core-collapse SNe, bringing gas temperatures beyond the peak of the cooling function (see also the discussion in \citealt{kobayashi2007}).
The impact of the IMF and other feedback models will be discussed in a forthcoming paper (Kobayashi 2020, in prep).

\subsection{Star formation, feedback, and chemical enrichment} 

Once star formation criteria of a gas particle is satisfied, we form a new star particle. Following the evolution of the star particles, we distribute thermal energy and mass of each chemical element to the surrounding gas particles \citep{kobayashi2004}. 
The neighbour particles are chosen to have approximately a fixed number, $N_{\text{FB}}=64$, and the energy and mass that each gas particle receives is weighted by the smoothing kernel \citep{kobayashi2007}. Hence, the feedback and chemical enrichment depend on the local gas density. When the cooling timescale is calculated, chemical compositions of neighbour particles are also used to include some effects of gas mixing in star formation \citep{haynes2019}\footnote{This is not included in our cosmological simulations such as in \citet{vincenzo2018a,vincenzo2018b}.}.

Following the method by \citet{kobayashi2004}, star particles  are treated as simple stellar populations (SSPs), hosting (unresolved) stars with the same ages and metallicities, and a spectrum of masses following the assumed IMF. In our simulation the adopted IMF is that of \citet{kroupa2008}, which has almost the same shape as that of \citet{chabrier2003}.
We also adopt the same metallicity-dependent stellar lifetimes as in \citet{kobayashi2004}.

\begin{figure}
\centering
\includegraphics[width=8.0cm]{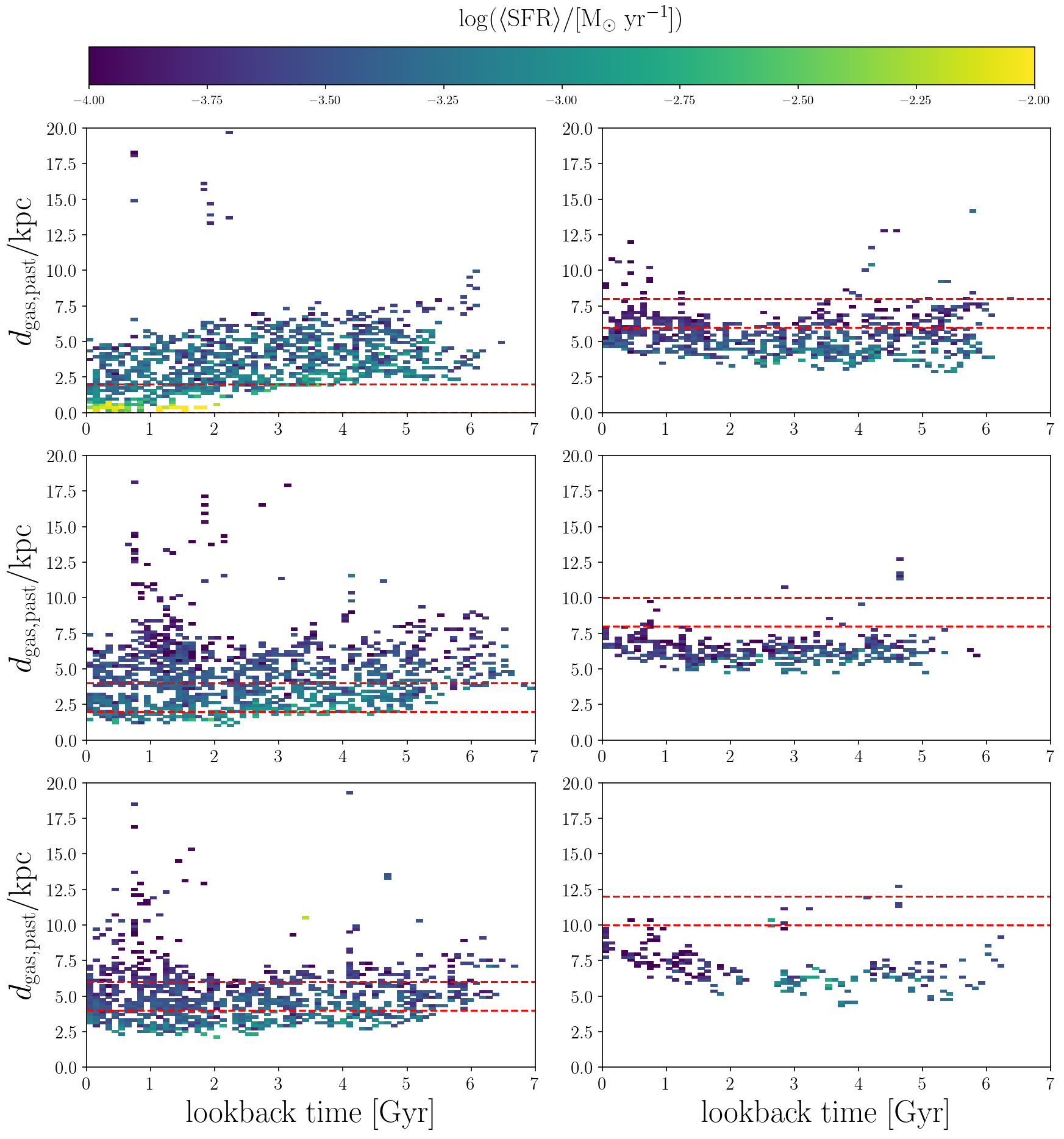} 
\caption{Same sequence of diagrams as those shown in the various columns of Fig. \ref{fig:gas-flow_thin-disc}, with the average SFR corresponding to the colour-coding. }
\label{fig:flow_SFR}
\end{figure}

Our chemodynamimcal simulation code includes {\it all} of major stellar nucleosynthetic sources in the cosmos: core-collapse  supernovae (Type II SNe and hypernovae, \citealt{kobayashi2006}), and Type Ia SNe \citep{kobayashi2009}, asymptotic giant branch stars (AGBs) \citep{kobayashi2011b}, and stellar winds from stars of all masses and metallicities.  
The feedback from the star formation activity 
depends on the metallicities and ages of the stellar populations. For the stellar yields and thermal energy feedback, 
we follow the same prescriptions as in \citet{kobayashi2011b}. Our simulation does not include failed SNe \citep{vincenzo2018a,vincenzo2018b,vincenzo2019c}, which are important to reproduce C/N and N/O in galaxies but not [Fe/H] and [$\alpha$/Fe] in this paper.

Type Ia SNe are the most important factor for predicting  [$\alpha$/Fe] ratios, and our progenitor model is the best constrained among other galaxy simulation codes.
Our model is based on the single-degenerate scenario of \citet{kobayashi2009}, where the deflagration of a Chandrasekhar-mass C+O white dwarf (WD) is triggered by the accretion of H-rich material from a main sequence or red giant companion star in a binary. During the accretion, it is assumed that metallicity-dependent WD winds can stabilise the mass-transfer for a wide range of binary parameters, allowing the system to give eventually rise to a Type Ia SN event (see, \citealt{kobayashi1998}, for more details). Therefore, our Type Ia SN rate depends on metallicity, and -- since the WD wind becomes very weak at [Fe/H]$<-1.1$ \citep{kobayashi1998} -- the assumed Type Ia SN rate is suppressed from star particles with [Fe/H]$<-1.1$. Without this effect, it is not possible to reproduce the observed [$\alpha$/Fe]--[Fe/H] relations in the Milky Way \citep{kobayashi2019}.

%%%%%%%%%%%%%%%%%%%%%%%%%%%%%%%%%%%%%%%%%%%%%%%%%%%%%%%%%
%%%%%%%%%%%%%%%%%%%%%%%%%%%%%%%%%%%%%%%%%%%%%%%%%%%%%%%%%
%%%%%%%%%%%%%%%%%%%%%%%%%%%%%%%%%%%%%%%%%%%%%%%%%%%%%%%%%
%%%%%%%%%%%%%%%%%%%%%%%%%%%%%%%%%%%%%%%%%%%%%%%%%%%%%%%%%
%%%%%%%%%%%%%%%%%%%%%%%%%%%%%%%%%%%%%%%%%%%%%%%%%%%%%%%%%
%%%%%%%%%%%%%%%%%%%%%%%%%%%%%%%%%%%%%%%%%%%%%%%%%%%%%%%%%
%%%%%%%%%%%%%%%%%%%%%%%%%%%%%%%%%%%%%%%%%%%%%%%%%%%%%%%%%
%%%%%%%%%%%%%%%%%%%%%%%%%%%%%%%%%%%%%%%%%%%%%%%%%%%%%%%%%

\begin{figure*}
\centering
\includegraphics[width=16.4cm]{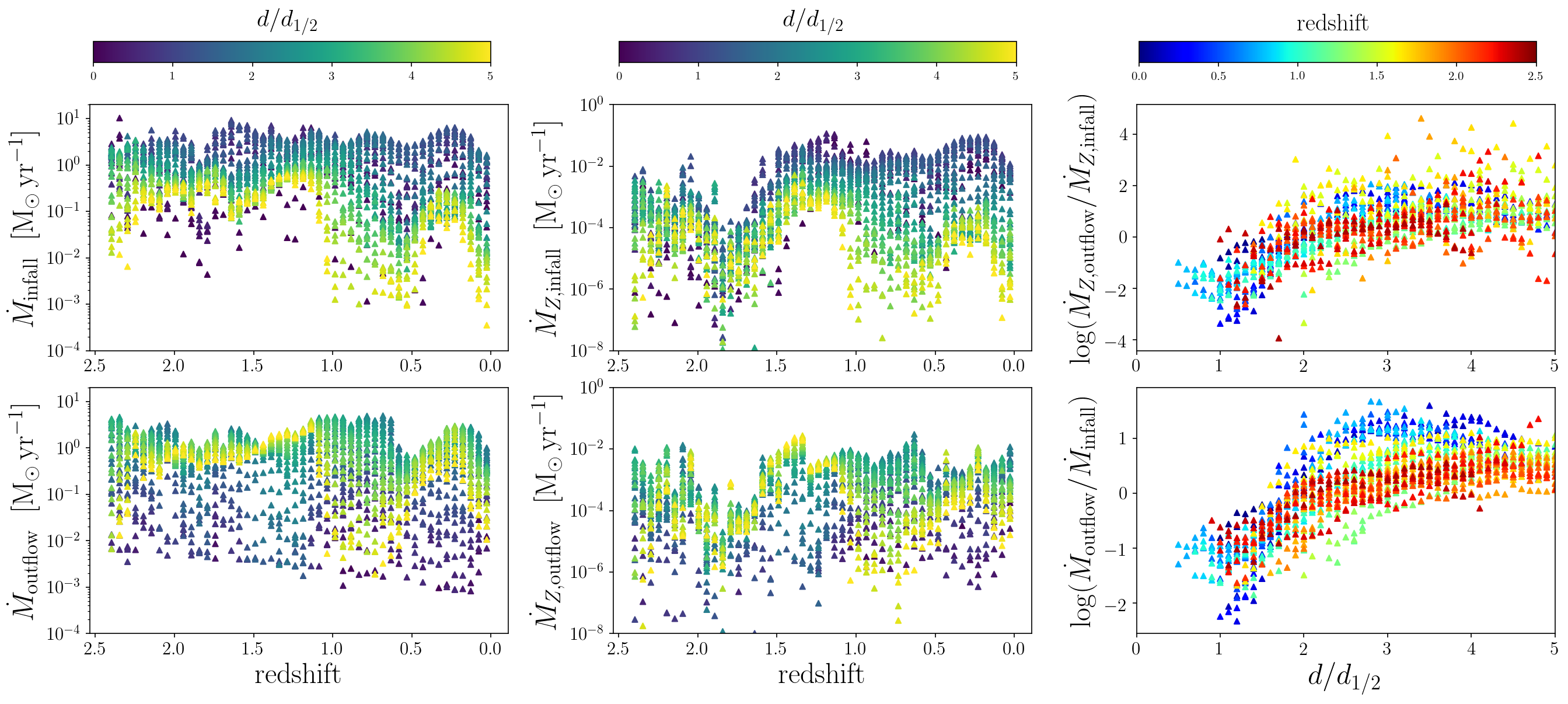} 
\caption{ \textit{ \textbf{(Left panels)}} The redshift evolution of the infall rate of gas (top panel) and outflow rate of gas (low panel) to/from our simulated disc; the colour-coding represents the galactocentric distance normalised to the half-mass radius, $d_{1/2}$, at each redshift. \textit{ \textbf{(Middle panels)}} The redshift evolution of the infall rate of metals (top panel) and outflow rate of metals (low panel); the colour-coding represents the galactocentric distance normalised to the half-mass radius at each redshift. \textit{ \textbf{(Right panels)}} The radial profile of the outflow-to-infall rate of metals (top panel) and gas (bottom) as a function of redshift, which is the colour-coding.
}
\label{fig:flow_redshift}
\end{figure*}

\section{Results} \label{sec:results}

Following the same procedure as in our previous works (e.g., \citealt{vincenzo2018b,vincenzo2019}), we define our simulated galaxy, by following the ID numbers of the gas particles that were in the galaxy at redshift $z$, in order to determine their coordinates at redshift $z+dz$; these coordinates define a particular region, $\mathcal{V}_{G}(z+dz)$, of the simulation volume. Then, at redshift $z+dz$, we fit the density distributions of \textit{all} the gas particles in $\mathcal{V}_{G}(z+dz)$, along the $x_{\text{gas}}$, $y_{\text{gas}}$, and $z_{\text{gas}}$ coordinates with three Gaussian functions. Therefore, the galaxy at redshift $z+dz$ is defined by considering all gas and star particles lying within $4\sigma$ of the three fitting Gaussian functions. This operation is repeated for all available snapshots, by starting from redshift $z=0$. 
Note that we do not apply bulge-disc-halo decomposition analysis, and our disc is defined only with the present location of stars and gas particles, by assuming a cut in the height $|h|<1\;\text{kpc}$ and galactocentric distance $d <14\;\mathrm{kpc}$; with this definition, the mass of the disc is $\log(M_{\star,\text{disc}}/\text{M}_{\sun}) = 10.91$.

 We find that at redshift $z=0$, the total stellar and gas masses of the simulated galaxy are $\log(M_{\star}/\text{M}_{\sun}) = 11.24$ and  $\log(M_{\text{gas}}/\text{M}_{\sun}) = 10.12$, respectively; moreover, the virial mass of the DM halo is $\log(M_{200}/\text{M}_{\sun}) = 12.23$. By applying a bulge-disc decomposition analysis, we find that the effective radius of the bulge and disc in the r-band are $r_{\text{bulge}}=1.63\;\text{kpc}$ and $r_{\text{disc}}=3.61\;\text{kpc}$, respectively. Finally, within the bulge effective radius, the mass of the bulge is $\log( M_{\star,\text{bulge}}/\text{M}_{\sun} ) = 10.66$.

 \subsection{The star formation history of the simulated disc}
 
 The age distribution of the stellar populations in the galaxy disc at $z=0$ is shown in Fig. \ref{fig1}(a); to make the histogram, we only consider stellar populations that have heights above or below the galactic plane $|h|<1\;\text{kpc}$ and galactocentric distances $d <14\;\mathrm{kpc}$. 
 In Fig. \ref{fig1}(b) the orange distribution corresponds to the normalised histogram considering all stellar populations in the galaxy disc with galactocentric distance $d_{\star} < 3\;\text{kpc}$, whereas the blue histogram includes only stars with $d_{\star} \ge 3\;\text{kpc}$. The three histograms in Fig. \ref{fig1} are all normalised to unity. We note that some of the stars with $d_{\star} \lesssim 1.6\;\text{kpc}$ in Fig. \ref{fig1} may be part of the bulge component.

The predicted galaxy SFH in Fig. \ref{fig1}(a) is characterised by several peaks at ages $>8$ Gyr; those ages correspond to the birth times of the halo and thick-disc stellar populations, which were accreted in majority from various mergers and accretion events. Then, in the last $\approx8$ Gyr, we witness at the formation of the majority of stars in the thin-disc component of the present-day galaxy, which is not altered by any major merger event. 
This is also consistent with another chemodynamical model in \citet{kobayashi2011} where 50 per cent of the stars in the Solar neighbourhood are younger than $8$ Gyr. Finally, from the histograms in Fig. \ref{fig1}(b), we find that $\approx68.7$ per cent of the stars with ages $>7$ Gyr are within $3\;\text{kpc}$ from the galaxy centre, with the remaining percentage lying in the outer disc, comprising mostly stars in the thick disc component with higher average galactic latitudes and velocity dispersion. This is qualitatively consistent with what is found in other simulations \citep{kobayashi2011,grand2018a,grand2020,fragkoudi2019b}.

In order to understand how the star formation activity propagated on the galaxy disc as a function of time, we divide the simulated gaseous disc in many concentric annuli and compute the evolution of the average star formation rate (SFR) within each annulus, starting from the present time and going back to $7$ Gyr ago; the results of our analysis are shown in Fig. \ref{fig2}(a). At first glance, the star formation activity seems to propagate from the inside out, with the average asymptotic SFR decreasing when moving outwards, at any fixed evolutionary time. This could be due to the growth of the disc. In the bottom panel (Fig. \ref{fig2}b), we show the same figure but normalized with $d_{1/2}$, which also shows the inside-out growth especially at $d/d_{1/2}>1.5$. Therefore, the inside-out growth can be characterised with two effects; (1) the SFR is always higher at the centre, which is due to the gas density and the exponential profile of the disc. (2) the onset of the disc formation is delayed, which appears at 6 Gyr for $d/d_{1/2}=1.5-2$ and 2 Gyr for $d/d_{1/2}=2-2.5$, respectively.
The half-mass radii of the simulated galaxy disc are: $d_{1/2} = 1.46, 1.38, 3.53, 4.43, 5.27$, and $6.05$ kpc at redshift $z=2, 1.5, 1, 0.5, 0.1$, and $0$, respectively.
These values are consistent with the observational findings of \citet{frankel2019}. 
Finally, we note that some of the star-forming gas particles at $d_{\text{gas}}=[0,2]\;\text{kpc}$ in Fig. \ref{fig2}(a) may form bulge stars later; however, we do not apply any additional selection for the gas particles. 

\subsection{The [$\alpha$/Fe]--[Fe/H] chemical abundance pattern}

An important observational feature that should be predicted by MW chemodynamical simulations is given by the dichotomy in [$\alpha$/Fe]--[Fe/H] at the Solar neighbourhood between thick- and thin-disc stars \citep{hayden2014,weinberg2019}. Our predictions for [$\alpha$/Fe]--[Fe/H] in the disc, by considering different bins of present-day galactocentric distances, are shown in the left panels of Fig. \ref{fig:alphafe-simulation}; the colour-coding in the figure corresponds to the average age of the stellar populations, whereas the dashed and solid red contours represent -- respectively -- the 10 and 20 per cent levels in the number of stars as normalised with respect to the maximum of the 2-D distribution. 

The predicted number density (red contours) of our simulation in Fig. \ref{fig:alphafe-simulation} are qualitatively compared with the [Mg/Fe]--[Fe/H] abundance patterns as observed by APOGEE-DR16 \citep{ahumada2019}, which are shown in the right panels of Fig. \ref{fig:alphafe-simulation}; the sample of MW stars from APOGEE-DR16 is obtained by adopting the same selection criteria that \citet{weinberg2019} followed for APOGEEE-DR14. We note here that the $\alpha$-element abundances shown in the left panels of Fig. \ref{fig:alphafe-simulation} for our simulation have been  obtained by empirically re-scaling our predicted O abundances to match the low-metallicity plateau of [Mg/Fe] as observed by APOGEE-DR16, which is at $\approx 0.35$ dex.

In the inner regions of our simulated galaxy there is an almost continuous trend in [$\alpha$/Fe] versus [Fe/H], which then breaks up into two different components when moving towards the outer annulii, where old/high-[$\alpha$/Fe] and young/low-[$\alpha$/Fe] populations become more and more separated with respect to each other, which corresponds to thin- and thick-disc populations identified in \citet{kobayashi2011}
(see also Fig. 1 of \citealt{kobayashi2016iau}). In particular, as we move towards the outer annulii, the low-[$\alpha$/Fe] thin-disc population is characterised by a metallicity distribution function (MDF) which is predicted to peak towards lower [Fe/H], in qualitative agreement with observations. 

The chemical composition of the star particles in our simulation is the same as that of the gas particles from which the stars were born. Therefore, the stellar MDF of the low-[$\alpha$/Fe] thin-disc population -- which mostly comprises young stars -- is almost the same as the MDF of the gas particles in the galaxy disc, which have -- on average -- their metallicity decreasing as a function of radius (see Section \ref{subsec:migration}, but also \citealt{kobayashi2011,vincenzo2018b}). This explains why, in Fig. \ref{fig:alphafe-simulation}, the low-[$\alpha$/Fe] thin-disc population moves towards higher [Fe/H] as we consider outer annulii. 

Note that the predicted dispersion of [$\alpha$/Fe] at fixed [Fe/H] seems larger than APOGEE, particularly in the innermost regions, which are -- however -- highly affected as well by dust extinction in the observations. Even so, compared with observations, our simulation might predict too many metal-rich stars with $\text{[Fe/H]}\approx 0.5$, which could require some modification of feedback modelling.  
Although APOGEE data do not reach the very low metallicities that can be seen in our simulation, particle-based simulations like ours tend to produce too many metal-poor stars at [Fe/H]$<-1$ (e.g., see Fig. 15 of \citealt{kobayashi2011}), whereas grid-based simulations tend to produce too narrow metallicity distribution functions (e.g., see \citealt{grand2018a}). 

In Fig. \ref{fig:apokasc} we compare the predicted age distribution of the stars in the [$\alpha$/Fe]--[Fe/H] diagram with the observed age distribution from the second APOKASC catalogue of \citet{pinsonneault2018}, which provides stellar ages for a sub-sample of $\approx 7000$ stars in common between APOGEE and Kepler. 
Fig \ref{fig:apokasc}(a) contains all stars in our simulated galaxy disc, namely with vertical height $|h_{\star}|<1\;\text{kpc}$ and galactocentric distance $d_{\star}<14\;\text{kpc}$, and the thick- and thin-disc sequences are highlighted by the red solid and dashed contours, representing the 10 and 20 per cent levels in the number of stars. Furthremore, we note that the chemical abundances in Fig. \ref{fig:apokasc}(b) are from APOGEE-DR16, whereas  \citet{pinsonneault2018} used chemical abundances from APOGEE-DR14. The comparison shows that our simulation can qualitatively capture the main observed age and chemical abundance distributions of the stars in the thick- and thin-disc sequences of the MW. 

If we look into the details, also in Fig. \ref{fig:apokasc}, there are some mismatches between observations and models, mostly due to systematic uncertainties in the simulation recipes for star formation and feedback as described above for the comparison in Fig. \ref{fig:alphafe-simulation}. Nevertheless, we remark on the fact that -- even though they provide a much more precise method than the classical isochrone fitting -- asteroseismic ages like those presented in \citet{pinsonneault2018} are also affected by some systematic uncertainties, since they make use of an empirical calibrations relating the stellar mass and radius with the frequency spacing, $\Delta \nu$, and the mode frequency of the normalised oscillation power spectrum, $\nu_{\text{max}}$, from the Kepler light curve (see also the detailed discussion in \citealt{lebreton2014} about the main systematic uncertainties coming from stellar evolution models, which are assumed in  the asteroseismic analysis).  

We also note that the thick- and thin-disc components are not defined in our analysis, but their stellar populations naturally emerge from the simulation itself; 
from an observational point of view, chemical abundances have been the best way to define, discriminate, and characterise those two components \citep{bensby2003,reddy2003}, without applying any particular definition in the data analysis but only earlier on in the development of the survey observation strategy; this explains why different spectroscopic surveys have different relative numbers of thick- and thin-disc stars (e.g., \citealt{bensby2014,magrini2018,feuillet2018,feuillet2019}).
In cosmological hydrodynamical simulations, a thick-disc component in simulated disc galaxy was firstly identified kinematically by \citet{brook2004}, and there are several other simulations that have found similar results like those presented here; among them, interesting studies are \citet{kobayashi2011,brook2012,mackereth2018,grand2018a,tissera2019,font2020}.
Different from these previous works, our simulations can well predict [$\alpha$/Fe], which made us possible to apply exactly the same approach as in the observational papers.
Our thin-disc population naturally comes out in the simulation with low-[$\alpha$/Fe] and ages $< 7$ Gyr.
In the following sections, we focus our analysis on the stellar population with ages younger than $7$ Gyr, referring to them as ``thin-disc'' stars.
Although there is a small number of young high-[$\alpha$/Fe] stars, we do not apply any selection with [$\alpha$/Fe] values.

\begin{figure}
\centering
\includegraphics[width=8.0cm]{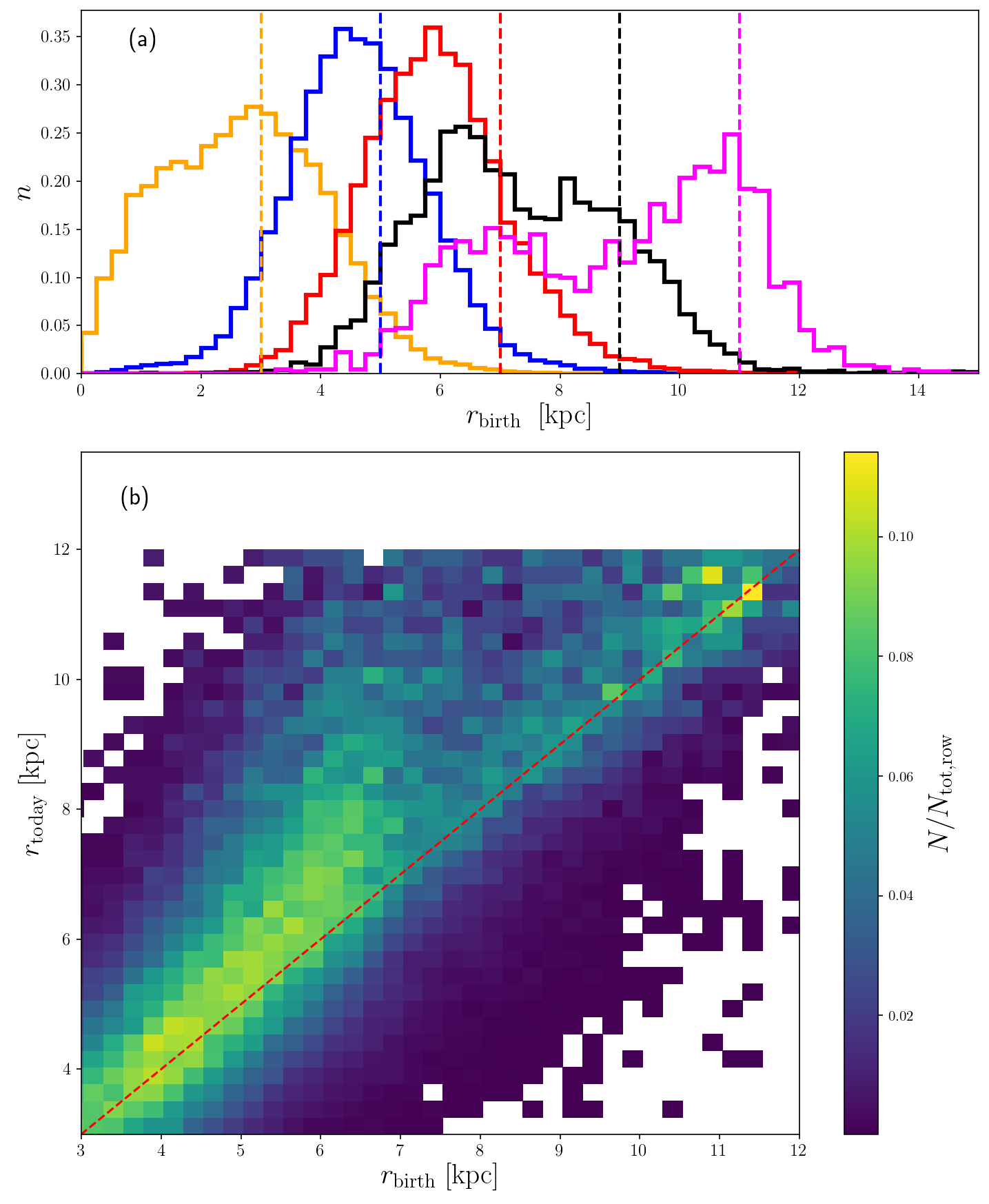} 
\caption{ \textit{ \textbf{(a)}} The distribution of the birth-radii of the thin-disc stars, $r_{\text{birth}}$, by considering different bins of present-day radii ($r_{\text{today}}$) with width $\Delta R = 2\;\text{kpc}$ (different colours in the figure); the vertical dashed lines with different colours correspond to the mean present-day radius of each distribution of birth-radii. \textit{ \textbf{(b)}} The distribution of the birth radii $r_{\text{birth}}$ compared to the present-day radii $r_{\text{today}}$ for the thin-disc disc stars in our simulated galaxy; the 2-D histogram is normalised along each row of present-day radii, and the colour-coding represents the normalised number of stars in each grid element. 
The red dashed line indicates no migration. }
\label{fig:migration}
\end{figure}

\begin{figure}
\centering
\includegraphics[width=8.0cm]{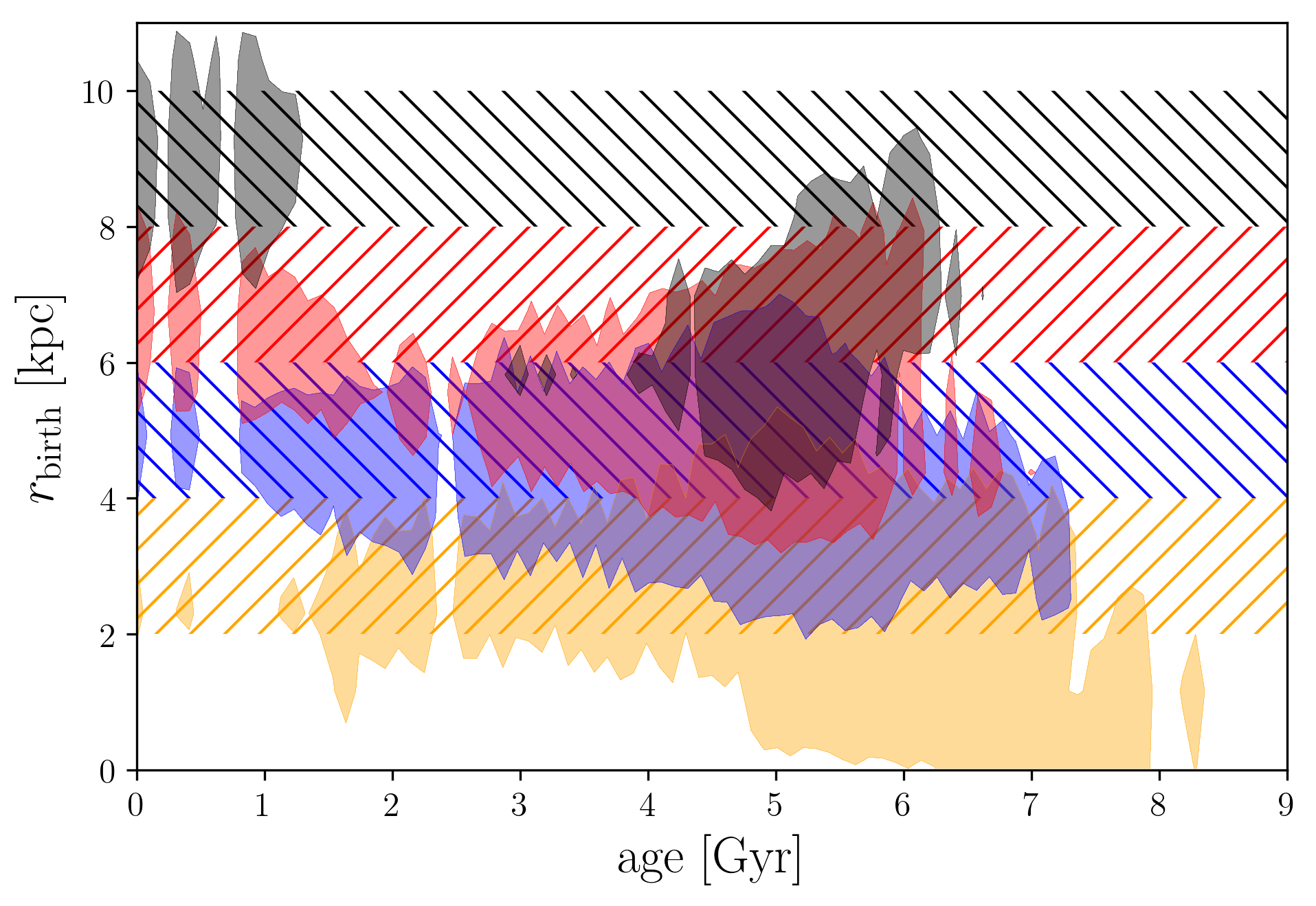} 
\caption{ The distribution of the birth-radii, $r_{\text{birth}}$, of the disc stars as a function of their age. Different colours correspond to different bins of present-day radii, indicated by the horizontal area of the same colour. For simplicity, the filled contours represent the $20$ per cent level in the number of stars, as normalised with respect to the maximum of each 2-D distribution }
\label{fig:migration-age}
\end{figure}

\subsection{Gas inflows}

The gas-phase chemical abundances that we observe in the thin disc at the present time are the result of a series of different physical and dynamical processes. Chemical abundances can depend on \textit{(i)} how the star formation activity in the galaxy varied as a function of radius and time (Fig. \ref{fig2}a), \textit{(ii)} radial flows of gas and metals in the disc, \textit{(iii)} accretion of gas and metals from the environment (infall), and \textit{(iv)} galactic winds (outflows).
In order to understand the impact of these gas flows on the chemical enrichment history of the galaxy, we follow the motions of gas particles by tracing their ID numbers in our chemodynamical simulation.

The gas inflow history that determined the formation and evolution of the present-day thin disc of our simulated galaxy is shown in Fig. \ref{fig:gas-flow_thin-disc}; we select all gas particles that are on the thin disc at $z=0$ and explore the evolution of their galactocentric distances as a function of look-back time. The colour-coding of each subgroup of figures (different columns) represents the following average quantities, moving from left to right: average stellar mass as computed for each grid element of the 2-D histogram, galactic height $|h_{\rm gas}|$, [Fe/H] abundance, and [$\alpha$/Fe] ratio. Finally, within each column, the different rows from top to bottom correspond to different bins of present-day galactocentric distances, as marked by the horizontal red dashed lines.

The formation and chemical evolution of the present-day simulated gaseous thin disc is determined by two main dynamical processes, both clearly visible in Fig. \ref{fig:gas-flow_thin-disc}: \textit{(i)} radial gas flow, which mostly involve star-forming and already chemically-enriched gas, moving along the galactic plane (i.e. at low $|h_{\text{gas}}|$), and \textit{(ii)} infall, i.e., accretion of non-star-forming, metal-poor and low [$\alpha$/Fe] gas. While approaching the galactic plane, the infall components gather the nucleosynthetic products of the core-collapse SNe exploding on the disc (more $\alpha$-elements than Fe), and therefore have their [$\alpha$/Fe] ratios increased just before they fall onto the disc. The fact that the radial gas flow component is star-forming whereas the infall component is non-star-forming is shown in Fig. \ref{fig:flow_SFR}. 
The radial flow component is dominant in the innermost galaxy regions, for galacticentric distances in the range $0 \lesssim d_{\text{gas}} \lesssim 5 \; \text{kpc} $, whereas the infall component is what determines the formation and chemical evolution of the thin disc in the outer galaxy regions, as defined in the range $d_{\text{gas}} \gtrsim 5 \; \text{kpc} $, particularly at $1$-$3$ Gyr ago.

In the radial flow component, as the gas particles cool and move on the galactic plane towards the inner galaxy regions, their [Fe/H] abundances steadily increase as a function of time mostly because of Type Ia SNe, which produce more iron than $\alpha$-elements; this also explains the decrease of [$\alpha$/Fe] in the radial gas flow component as a function of time, as the gas particles flow along the galactic plane. 

By fitting with a linear law the relation between the galactocentric distance and the look-back time for the radial flow component in Fig. \ref{fig:gas-flow_thin-disc}, the average velocity of the radial gas flows is estimated to be $0.7 \pm 0.14$ km/s, by considering disc gas particles with $0 < r_{\text{today}} < 4\,\text{kpc}$; this value is consistent, for example, with the findings of previous works \citep[e.g.,][]{laceyfall1985,bilitewski2012} and the typical assumptions of chemical evolution models \citep[e.g.,][]{portinari2000,spitoni2011,mott2013,grisoni2018}. Note that unlike these classical chemical evolution models, our radial flow velocity is not set as a parameter but is calculated as a consequence of dynamical evolution from cosmological initial conditions.

\begin{figure}
\centering
\includegraphics[width=8.0cm]{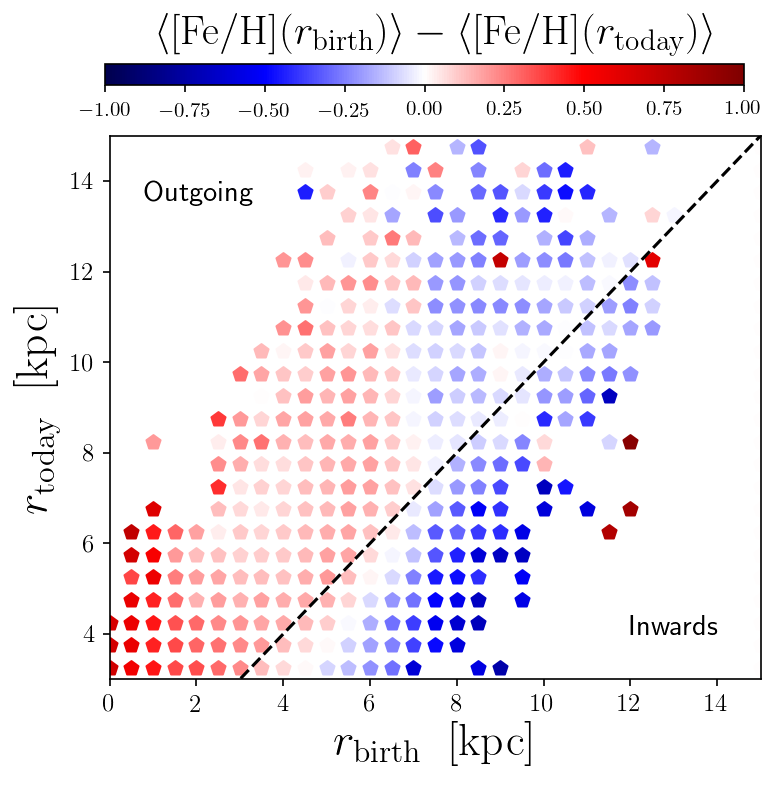} 
\caption{ The difference between the average [Fe/H] at the birth-radius, $\langle \text{[Fe/H]}(r_{\text{birth}}) \rangle$, and the average [Fe/H] at the present-day radius, $\langle \text{[Fe/H]}(r_{\text{today}}) \rangle$, in the $r_\text{today}$--$r_{\text{birth}}$ diagram for the thin-disc stars. The dashed line indicates no migration.
}
\label{fig:dfe-dv-dr_migration}
\end{figure}

\begin{figure}
\centering
\includegraphics[width=8.0cm]{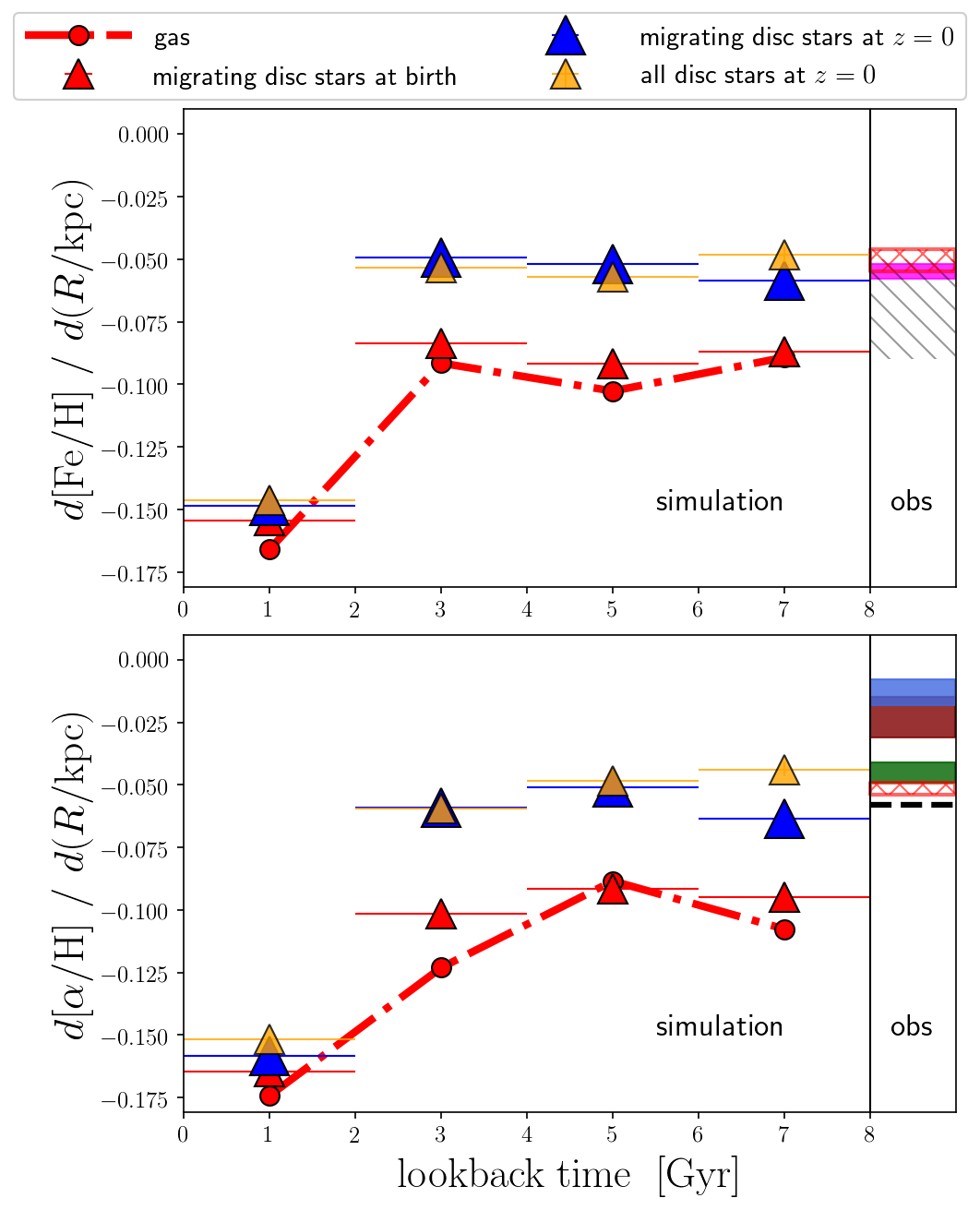} 
\caption{ \textbf{\textit{(Left-hand side)}} The evolution of the slope of the predicted [Fe/H] (top panel) and [$\alpha$/H] (bottom panel) radial gradients in the simulated disc. The red dash-dotted curve which connects the filled red circles corresponds to the gas-phase slope; the red triangles correspond to the stellar abundance gradients, when using the birth-radii of the stars instead of their present-day radii; the yellow triangles represent the present-day stellar abundance gradients; and blue triangles represent-day stellar abundance gradients in the migrating stellar populations on the disc. \textbf{\textit{(Right-hand side)}} Observational data for the slope of the [Fe/H] (top panel) and [$\alpha$/H] (bottom panel) gradient, by using Cepheids (\citealt{luck2011}, in magenta; \citealt{korotin2014}, black dashed line; and \citealt{genovali2015}, in green); young and old Planetary Nebulae, in brown and blue, respectively (data from \citealt{stanghellini2018}); young open clusters (\citealt{magrini2017}, shaded area with grey diagonal lines); and all disc stars with Galactocentric distances from $d = 5$ to $10\;\text{kpc}$ with $|z| < 0.5 \; \text{kpc}$ from APOGEE-DR16 (\citealt{ahumada2019}, red area).   }
\label{fig:feh-slope}
\end{figure}

\subsection{Gas outflows}

Due to supernova feedback, some of gas is heated and ejected from the galaxy disc, carrying heavy elements produced by disc stars.
This metal-loss by winds is extremely important for understanding the metallicities in circumgalactic and intergalatic medium \citep[e.g.,][]{RenziniAndreon14}, as well as the mass--metallicity relation of galaxies \citep[e.g.,][]{kobayashi2007}.
Also in our MW simulation, there are gas particles that are not on the disc at present, but have been on the disc in the past -- they represent the outflow component.

We recall there that in Fig. \ref{fig2} we show the inside-out growth on the simulated disc; similarly, infall and outflow depend on the galactocentric distance.
In Fig. \ref{fig:flow_redshift} (right panels), we show the ratios of the outflow and infall rates of gas and gas metals as a function of the galactocentric distance, from $z=0$ to $z=2.5$. The outflow component is computed by selecting the $x_{\text{gas}}$, $y_{\text{gas}}$, and $z_{\text{gas}}$ coordinates (together with their mass, density, and metallicity) of all the gas particles that were on the galactic disc at redshift $z+dz$ and have been expelled between $z+dz$ and $z$; the inflow component consists of all the gas particles that were in the circumgalactic medium at $z+dz$ and have been accreted between $z+dz$ and $z$. Then, at each available redshift, we divide the galaxy disc in many concentric annulii with length $\Delta R = 0.1\;\text{kpc}$ and compute the total amount of gas and metals in the gas-phase that are accreted and expelled per unit time.
Note that radial gas flow is not included in this figure, while the outflow rate includes both winds and fountains (i.e., some of the ejected gas particles may fall back).

Since our simulated galaxy disc grows -- on average -- in mass and size as a function of time (see also \citealt{vincenzo2019}), we normalise the galactocentric distance to the half-mass radius, $d_{1/2}$, of the galaxy at each redshift. 
Although there is a significant scatter, we find that the average profile of the ratio between the outflow and infall rates of metals follows a time independent radial profile in this halo; innermost regions of the galaxy are always infall-driven, while less bound outer regions are always outflow-driven showing higher metal outflow-to-infall ratios than in innermost regions.
The redshift-independence of that radial profile stems from the growth of the dark-matter halo as a function of time, which -- on the one hand -- determines the retention of metals expelled by stellar winds and SN explosions as a function of radius, and -- on the other hand -- regulates the amount of gas and metals accreted from the environment. The outer galaxy regions have a lower metal retention efficiency than the inner galaxy regions; in particular, the outer galaxy regions have more metals being expelled through outflows than metals being accreted. Conversely, in the inner regions, metals are preferentially retained because of the larger gravitational energy.

If the redshift-independent radial profile of outflow-to-infall ratios also exists in other MW-type halos, then there should be always a negative radial gas-phase metallicity gradient, which namely decreases as a function of the galactocentric distance. 
The negative metallicity gradient is primarily developed by the fact that there is a radial gradient in the SFR (see the asymptotic values of the SFR at different radial intervals in Fig. \ref{fig2}).
The gradient is maintained by the fact that -- at any redshift -- the metal outflow rate becomes increasingly larger than the metal infall rate when moving outwards.
Inner regions always increase metals, while outer regions loose metals.
The evolution of [Fe/H] and [O/H] of each region are more complicated as they are also affected by the infall and outflow of hydrogen, which is discussed in the next Section.

In Fig. \ref{fig:flow_redshift}, we also show how the infall rate of gas and metals (top left and top middle, respectively) and the outflow rate of gas and metals (bottom left and bottom middle, respectively) evolve as a function of redshift, with the colour-coding representing the galactocentric distance normalised to the half-mass radius at any given redshift. First of all, we notice that both infall and outflow rates have an oscillating evolution as a function of time, especially in the outer galaxy regions.
Secondly, gas infall takes place inside-out having higher rates at innermost regions, while outflow undergoes  outside-in at $z\gtsim1$ with higher rates at outer regions.

\subsection{Stellar migrations} \label{subsec:migration}

\begin{figure*}
\centering
\includegraphics[width=16.4cm]{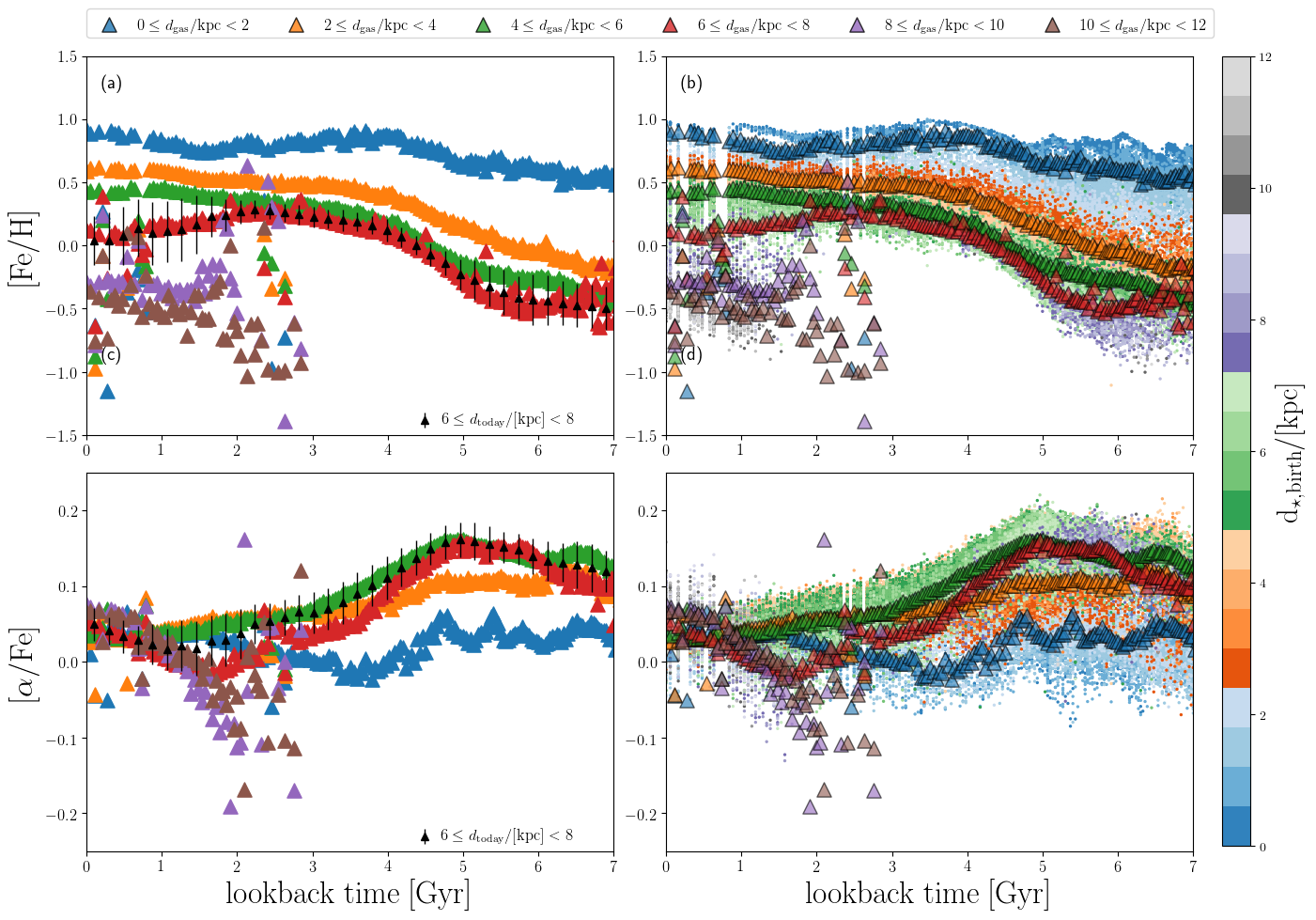} 
\caption{ \textit{ \textbf{(Left panels)}}: the gas-phase predicted [Fe/H] (top panel) and [$\alpha$/Fe] (bottom panel) versus \textit{look-back time} relations at different galactocentric annulii, $d_{\text{gas}}$ (different colours in the figure), compared with the predicted relations in the disc  stars at the present-day annulus $6 \le d_{\star,\text{today}} \le 8 \;\text{kpc}$ (black triangles with error bars). \textit{ \textbf{(Right panels)}}: same as in the left panels, but we also show the relations that we obtain when we put together all thin-disc stars and use their birth-radii, $d_{\star,\text{birth}}$, for the colour-coding. }
\label{fig:ofe-feh-age-migration}
\end{figure*}

Not only gas but also stars move during galaxy evolution, which causes extra mixing of metals. In our chemodynamical simulation, we follow the motions of star particles by tracing their ID numbers. 
The orbits of stars in a galaxy can be largely disturbed by various physical processes such as \textit{(i)} galaxy mergers, \textit{(ii)} satellite accretions, and \textit{(iii)} stellar migrations.
The impact on chemical abundances is an important question;
\citet{kobayashi2004} showed that galaxy mergers flatten radial metallicity gradients, while \citet{kobayashi2014} showed that a half of thick disc stars come from satellite accretion and have a different chemical abundances from the stars formed in-situ. 

Stellar migrations can, therefore, play an important role in the chemical evolution of galaxies \citep[e.g.,][]{schoenrich2009,minchev2011,minchev2018}.
Naively, the main physical mechanisms behind stellar migrations are churning and blurring \citep[e.g., ][]{selwood2002,schoenrich2009}, as well the overlap of the spiral and bar resonances on the disc \citep{minchev2011}, however, in this paper we consider any change in radius as migration as in \citet{kobayashi2014}. 

Stellar migrations can also represent a contamination process in which stars from a distribution of different birth-radii (having their own star formation and chemical-evolution histories, possibly affected by stellar migration itself) populate a different galaxy region at the epoch of their observation. It is important to quantify this mechanism of radial mixing of stars, particularly if we want to understand how the radial metallicity gradient in galaxies evolves as a function of time, by looking at the present-day chemical abundances in stellar populations with different ages.

In our simulation stellar migration is analysed by retrieving the birth-radii of all the stellar populations that are on the simulated galaxy disc at the present-time, and that formed on the disc in the past ($|h|<1$ kpc); all of these stars have ages $\lesssim 7\;\text{Gyr}$, since stars with older ages have mostly been accreted from mergers and tidal interactions with satellites. There are also stars with age $<7\;\text{Gyr}$, which are on the disc today but were accreted in the past from other systems; these stars are not taken into account when we analyse the impact of stellar migration.

The results of our analysis of stellar migrations are shown in Figure \ref{fig:migration}, in which we only consider stars that have $|h| < 1 \;\text{kpc}$ at the present time. In the top panel, we show how the distribution of the birth radii, $r_{\text{birth}}$, varies when considering stars that reside at the present time in different galaxy annuli, from $r_{\text{today}}=2\;\text{kpc}$ to $r_{\text{today}}=12\;\text{kpc}$, with the annuli having width $\Delta r = 2\;\text{kpc}$ (different colours in the figure). 
In the top panel, we show how the stars in our simulation distribute in the $r_{\text{birth}}$ versus $r_{\text{today}}$ diagram, with the 2-D histogram being normalised by row, namely over the distribution of $r_{\text{birth}}$ for each bin of $r_{\text{today}}$.
The red dashed line indicates no migration, and there are a few stars following this line.
A larger fraction of stars are found above this line, meaning these stars move outwards. On the one hand, we find that approximately $66$ per cent of the stars which live on the galaxy disc at the present time have migrated outwards (namely, $d_{\star,\text{today}} - d_{\star,\text{birth}} > 0$), $29$ per cent have migrated outwards by at least $1\;\text{kpc}$ (namely, $d_{\star,\text{today}} - d_{\star,\text{birth}} > 1\;\text{kpc}$), $8$ per cent have migrated outwards by at least $2\;\text{kpc}$, and  $2$ per cent have migrated outwards by at least $3\;\text{kpc}$. On the other hand, we find that approximately $9$ per cent of the stars on the present-day disc have migrated inwards by at least $1$ kpc (namely $d_{\star,\text{birth}} - d_{\star,\text{today}} > 1\;\text{kpc}$), and $2$ per cent have migrated inwards by at least $2$ kpc. 

The bottom panel of Fig. \ref{fig:migration} shows the same results but with a different presentation to compare with Figure 3 of \citet{martinez-medina2016}.
The vertical dashed lines with different colours represent the mean values of $r_{\text{today}}$ for each distribution of $r_{\text{birth}}$. 
In their N-body simulations assuming a static DM potential, \citet{martinez-medina2016} find that the distribution of $r_{\text{birth}}$ is almost Gaussian for stars with given $r_{\text{today}}$, meaning that stars randomly move both inwards and outwards. We find, however, that stellar migration mostly involves stars migrating outwards, which is in agreement with the results of previous studies making also use of cosmological chemodynamical simulations (e.g., \citealt{brook2012,bird2013}). 

To understand the typical timescales over which stellar migration takes place, in Fig. \ref{fig:migration-age} we show our predictions for $r_{\text{birth}}$ versus stellar age, by considering different present-day annulii, which are marked by the horizontal areas of different colours. The filled contours represent the $20$ per cent level in the number of stars, as normalised with respect to the maximum of each 2-D distribution.  We find that the amount of stellar migration correlates well with the stellar age, in the sense that older stellar populations are also those that migrated more, as already shown -- for example -- in \citet{brook2012}. Finally, the typical timescale for the outgoing stellar migration is $\tau_{\text{migr}}\approx5.4$, $4.6$, $4.5$, and $4.9$ Gyr for thin-disc stars that -- at the present-time -- reside in the annuli centered at $3$, $5$, $7$, and $9$ kpc, respectively. These values are measured by fitting the predicted $\Delta r_{\text{migr}} = r_{\text{birth}} - r_{\text{today}}$ versus \textit{age} relations of the
stellar populations migrating outwards with a function of the
form $A - \exp{(\textit{age}_{\star}/\tau_{\text{migr}})}$.

In Fig. \ref{fig:dfe-dv-dr_migration}, we characterise the impact of stellar migrations in mixing stars of different metallicity on the simulated galaxy disc, by looking at its effect in the $r_{\text{today}}$ versus $r_{\text{birth}}$ diagram. To make the figure, we firstly bin all the migrating stellar populations according to their birth-radii and present-day radii; then we compute the average [Fe/H] for each bin of birth-radii and present-day radii as obtained from the previous binning procedure. In particular, the quantity $\langle [\text{Fe/H}](r_{\text{birth}}) \rangle$ represents the average [Fe/H] of thin-disc stars by considering their birth-radius, $r_{\text{birth}}$, whereas the quantity $\langle [\text{Fe/H}](r_{\text{today}}) \rangle$ corresponds to the average iron abundance at $r_{\text{today}}$, by considering all the stellar populations that at the present-time reside at $r_{\text{today}}$. Therefore, $\langle [\text{Fe/H}](r_{\text{today}}) \rangle$ is computed by including both the stars that were born in the disc and then migrated and the stars that were born in accreted stellar systems and reside on the disc at the present time (which represent a minor component). This way we can effectively quantify the effect of stellar migrations on the radial stellar metallicity gradient. 

The difference $\langle [\text{Fe/H}](r_{\text{birth}}) \rangle-\langle [\text{Fe/H}](r_{\text{today}}) \rangle$ in Fig. \ref{fig:dfe-dv-dr_migration} is the effect of migration on the average iron abundance of the stars on the disc.
This difference is positive at $r_{\rm birth} \ltsim 6$ kpc, where some of the stars move outwards to increase the average metallicity at $r_{\rm today}$.
Therefore, if we consider stellar populations that have migrated outwards, stellar migration mostly involves metal-rich stars that have moved towards more metal-poor regions at the present-time, even though -- in the outer disc -- there is a minor component of old migrating stellar populations that have migrated outwards and have lower [Fe/H] than in their present-day galactocentric annulus.
Conversely, the difference is negative for stars with $r_{\rm birth} > r_{\rm today}$, which means that most of the stars that migrated inwards are more metal-poor than the average metallicity at their present-day radius. 

\begin{figure*}
\centering
\includegraphics[width=16.4cm]{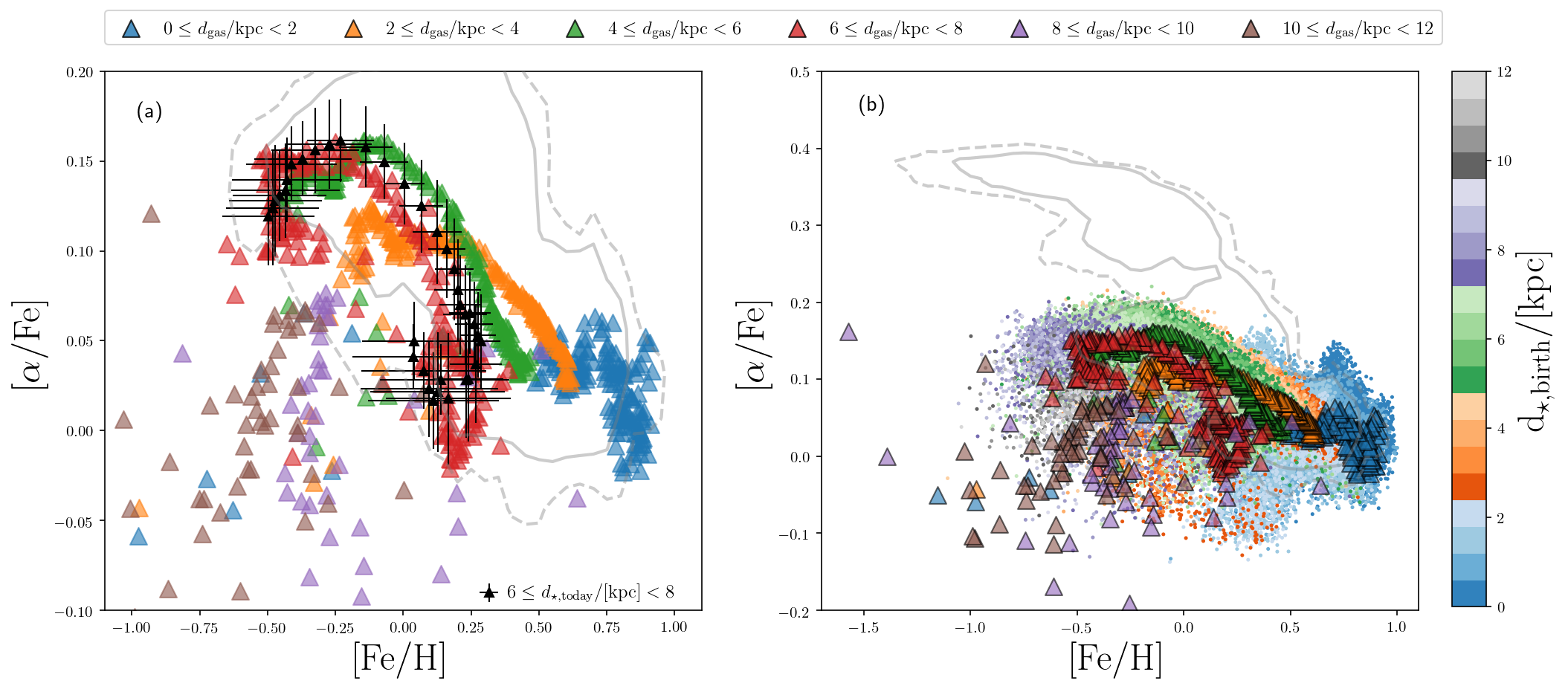} 
\caption{Same as in Fig. \ref{fig:ofe-feh-age-migration} but showing the predicted relations in the [$\alpha$/Fe]--[Fe/H] diagram for stars with ages $<8\;\text{Gyr}$. }
\label{fig:ofe-feh-age-migration2}
\end{figure*}

As shown in Fig. \ref{fig:migration}, the number of stars that migrated outwards is larger than the number of stars that migrated inwards; hence, considering what we also find in Fig. \ref{fig:dfe-dv-dr_migration},  stellar migrations should cause a flattening of the radial stellar metallicity gradient. This is demonstrated in Figure \ref{fig:feh-slope}, where we show how the slope of the radial [Fe/H] and [O/H] gradients 
evolve as a function of the stellar age, by considering all stellar populations existing on the disc at the present-time, including the stars accreted which were born off plane (dark yellow triangles).
The gradient evolution is almost the same even if we
consider only the stars that were formed in the simulated disc and are present in the disc today (blue triangles).
Then, we show how the radial metallicity gradient changes if we use the birth-radius of the stars instead of their present-day radius (red triangles). 
The overall effect of stellar migrations is to flatten the slope of the radial metallicity gradient sampled by stars with age $\gtrsim 2\;\text{Gyr}$ by $\approx 0.05\;\text{dex}\;\text{kpc}^{-1}$.

Finally, the red-dotted curve with filled red circles shows the evolution of the slope of the gas-phase metallicity gradient at the epoch corresponding to the mean stellar age of each bin, which exactly follows the red triangles.
If we exclude the effect of migrations (red triangles), the metallicity of the stars at a given age and radius should be very similar to the metallicity of the gas at the same radius and the corresponding redshift, and hence
the stellar gradient follows the evolution of the gas-phase gradients (red circles).

The gas-phase (and stellar) metallicity gradient becomes steeper 
from $2$ Gyr ago (redshift $z=0.25$) to the present ($z=0$); this is due to recent infall of metal-poor gas at
the solar neighbourhood and larger radii, which triggered the
formation of stars with relatively lower average metallicity at those
radii (see Fig. \ref{fig:gas-flow_thin-disc}).
This could be tested with on-going survey with integral field units of disc galaxies.

In Fig. \ref{fig:feh-slope} we also show  observations, most of which agree very well with the slopes that we predict for the radial [O/H] and [Fe/H] gradients for stars with age $>2$ Gyr, which correspond to $\approx 95$ per cent of all stars in the simulated disc (see also Fig. \ref{fig1}). In particular, we measure the slopes of the radial [Mg/H] and [Fe/H] gradients from APOGEE-DR16 \citep{ahumada2019}, by considering a sample of MW stars with signal-to-noise $\text{S/N}>80$, Galactic altitudes $|h|<0.5\;\text{kpc}$, and in the range of Galactocentric distances $5\leq R \leq 10\;\text{kpc}$, with the errors in the plot taking into account how the slope changes if we consider a minimum distance of $3\;\text{kpc}$. 
The slope that we measure for [Fe/H] with APOGEE-DR16 is in the range between $-0.046$ and $- 0.055$ dex/kpc, which is very similar to that for [Mg/H] (which is in the range between $-0.049$ and $-0.054$ dex/kpc) and [O/H] (which is in the range between $-0.047$ and $-0.054$ dex/kpc). 

The slope that we measure for [Fe/H], [Mg/H] and [O/H] in APOGEE-DR16 is consistent with that measured by \citet{genovali2015} for [Mg/H] in a sample of Cepheids in the MW thin disc ($-0.045\pm 0.004$ dex/kpc), and is of the same order of magnitude as that determined by other Cepheid observations, like \citet{luck2011}, who measure a slope of $-0.055 \pm 0.003$ kpc/dex for [Fe/H], and \citet{korotin2014}, who measure $-0.058$ dex/kpc for [O/H]. Finally, in Fig. \ref{fig:feh-slope}, we also show the slope that we derive for [Fe/H] from the sample of \citet{magrini2017} of young open clusters with ages $<2\;\text{Gyr}$, finding $-0.07 \pm 0.02$ kpc/dex. 

On the one hand, our predictions for the the slope of the radial metallicity gradients agree well with the APOGEE-DR16 observations. On the other hand, the predicted radial metallicity gradients in the gas-phase and in the youngest stellar populations at redshift $z=0$ are steeper than those observed in Cepheids \citep{luck2011,korotin2014} and young open clusters \citep{magrini2017}; as already mentioned above, this is due to a recent accretion event of metal-poor gas in the outer regions of the simulated galaxy disc, happening in the last $2$ Gyr (see Fig. \ref{fig:gas-flow_thin-disc}), that triggered the formation of a population of young stars in the outer regions that have relatively low average metallicities. 

Finally, we do not reproduce the slope of the [O/H] gradient as derived with planetary nebulae by \citet{stanghellini2018} for a sample of young (YPPNe; ages $\lesssim 7\;\text{Gyr}$) and old (OPPNe; ages $\gtrsim 7\;\text{Gyr}$) planetary nebulae in the Galaxy disc, with their ages being determined by comparing their observed C/N ratios with that predicted by post-AGB stellar evolution models of different mass.

\subsection{The impact of stellar migrations on the age--metallicity relation and on the thin-disc low-[$\alpha/\text{Fe}$] sequence}

In Figs. \ref{fig:ofe-feh-age-migration} and \ref{fig:ofe-feh-age-migration2} we explore the impact of stellar migrations on the predicted [Fe/H] and [$\alpha$/Fe] versus look-back time relations; in particular, the temporal evolution of the average [Fe/H] and [$\alpha$/Fe] in the ISM at different annulii is compared with the [Fe/H] and [$\alpha$/Fe] versus \textit{age} relations as predicted when putting together all thin-disc stars and using their birth-radii for the colour-coding. We also show the predicted [Fe/H] and [O/Fe] versus \textit{age} relations as predicted for the stars that live today in the galactic annulus as given by $6 \le d_{\star,\text{today}} \le 8 \; \text{kpc}$ (black triangles with error bars).

When we put together all the stars on the simulated galaxy disc, the predicted spread in the [Fe/H]--age and [$\alpha$/Fe]--age diagrams seems to be a consequence of the gas-phase chemical evolution tracks, taking place locally at the different birth-radii of the stars (see Fig. \ref{fig:ofe-feh-age-migration}); in fact, the temporal evolution of the gas-phase [Fe/H] abundances at different annuli closely follow the stellar [Fe/H] at the same birth time and radius of the stars. For example, if we select stars that today lie in the range $6 \le d_{\star,{\rm today}} \le 8 \; \text{kpc}$, their predicted average [Fe/H]--age relation (black triangles) closely traces the average gas-phase [Fe/H]--look-back time relation in the same annulus (red triangles), with the amount of dispersion given by the error bars following the gas-phase [Fe/H] in the inner annulus (as defined by $4 \le d_{\rm gas} \le 6 \; \text{kpc}$, green triangles). 

The gas-phase chemical evolution tracks, i.e., [Fe/H] versus lookback time, in Fig. \ref{fig:ofe-feh-age-migration} shift -- on average -- towards higher [Fe/H] as we consider inner galaxy regions. Interestingly, at approximately the solar neighbourhood (and beyond towards the outer annulii), [Fe/H] slightly decreases and then remains almost constant from $\approx7\;\text{Gyr}$ to $\approx5\;\text{Gyr}$ ago, whereas [$\alpha$/Fe] increases with time during the same time interval; a similar trend is also found in the last $\approx 2\,\text{Gyr}$. This is due to the aforementioned accretion of metal-poor gas, triggering star-formation activity on the galaxy disc, which causes [Fe/H] to decrease or remain constant while [$\alpha$/Fe] increases. 
These signatures of gas accretions that we see in Fig. \ref{fig:ofe-feh-age-migration} about $0$-$2$ and $5$-$7$ Gyr ago give rise to a characteristic feature in the predicted [$\alpha$/Fe]--[Fe/H] diagram in Fig. \ref{fig:ofe-feh-age-migration2}, for the solar neighbourhood (red triangles), respectively at $[\alpha/\text{Fe}] \approx 0$ and $[\text{Fe/H}] \approx 0$ and at $[\alpha/\text{Fe}] \approx 0.1$ and $[\text{Fe/H}] \approx -0.5$.

We find that the effect of stellar migrations can appear more in the predicted $[\alpha/\text{Fe}]$--$\text{age}$ relation than in the $\text{[Fe/H]}$--\textit{age} relation at the solar annulus, as follows. If we consider the predicted [Fe/H]--\textit{age} relation, the [Fe/H] abundances of the stars at the solar neighbourhood are well synchronised with the gas-phase [Fe/H] abundances at the same radius. However, there is a mismatch for $[\alpha/\text{Fe}]$ at $1$-$4$ Gyr (between black and red triangles), which is caused by migration. As we have shown in Fig. \ref{fig:migration-age} the timescale of migration is $\approx 5 \; \text{Gyr}$ and stars move $\pm 2 \; \text{kpc}$ at most. Therefore, at lookback times $t<1\;\text{Gyr}$, there is not enough time to have migration, whereas at $t>4\;\text{Gyr}$ we cannot see the migration effect because there is no difference in the abundance ratios at $4$-$6$ kpc and $6$-$8$ kpc. We note that these results are also more evident for the solar neighbourhood, where migration has a stronger effect in our simulation (see Fig. \ref{fig:migration}).

Finally, by looking at Figs. \ref{fig:ofe-feh-age-migration}(b) and \ref{fig:ofe-feh-age-migration2}(b), we note that there is a strong correlation between the birth radius and the position of the thin-disc stars in [$\alpha$/Fe]--[Fe/H] (small dots); in particular, stars that were born in the inner simulated galaxy regions tend to appear at higher [Fe/H] and lower [$\alpha$/Fe] than stars that were born in the outer annulii; this is our predicted effect of stellar migrations, and can be tested with observations. In particular, if we consider the stellar populations that formed in the inner galaxy ($0 < d_{\star,\text{birth}} < 2$), they were born from gas-phase abundances affected by the earliest stages of galaxy evolution, when chemical enrichment timescales were shorter and the SFR was higher than in the last $7$ Gyr; this chemical-evolution mechanism caused a faster evolution of [Fe/H] and hence lower average values of $[\alpha/\text{Fe}]$ later on in the gas-phase than in the outer annuli. These results seem to be in line with recent observational findings by \citet{ciuca2020}.

%%%%%%%%%%%%%%%%%%%%%%%%%%%%%%%%%%%%%%%%%%%%%%%%%%%%%%%%%
%%%%%%%%%%%%%%%%%%%%%%%%%%%%%%%%%%%%%%%%%%%%%%%%%%%%%%%%%
%%%%%%%%%%%%%%%%%%%%%%%%%%%%%%%%%%%%%%%%%%%%%%%%%%%%%%%%%
%%%%%%%%%%%%%%%%%%%%%%%%%%%%%%%%%%%%%%%%%%%%%%%%%%%%%%%%%
%%%%%%%%%%%%%%%%%%%%%%%%%%%%%%%%%%%%%%%%%%%%%%%%%%%%%%%%%
%%%%%%%%%%%%%%%%%%%%%%%%%%%%%%%%%%%%%%%%%%%%%%%%%%%%%%%%%
%%%%%%%%%%%%%%%%%%%%%%%%%%%%%%%%%%%%%%%%%%%%%%%%%%%%%%%%%
%%%%%%%%%%%%%%%%%%%%%%%%%%%%%%%%%%%%%%%%%%%%%%%%%%%%%%%%%

\section{Conclusions} \label{sec:conclusions}

In this paper we have analysed our self-consistent cosmological chemo-dynamical simulation of a MW-type galaxy in order to understand the roles of metal flows in galaxy formation and evolution.
Our primary purpose is to characterise the main physical and dynamical processes that determined the chemical evolution of the stellar and gaseous thin disc of a MW-type galaxy as a function of time and radius. 

In our simulated disc, the inner parts are mainly comprised by old stellar populations, although old stars can also be found at the solar neighbourhood (Fig. \ref{fig1}).
 Moreover, our simulation can qualitatively reproduce the observed radial variations of the bimodal distribution in [$\alpha$/Fe]--[Fe/H] from APOGEE-DR16 (Fig. \ref{fig:alphafe-simulation}), which corresponds to thick- and thin-disc stars. In particular, our low-$\alpha$ population tends to have younger ages than the high-$\alpha$ population (Fig. \ref{fig:apokasc}), and the location of the young low-$\alpha$ population moves in [$\alpha$/Fe]--[Fe/H], having lower [Fe/H] at the outer disc; finally, if we look at the innermost regions of the galaxy disc, there is an almost continuous trend in [$\alpha$/Fe]--[Fe/H]. These results are qualitatively in agreement with APOGEE observations \citep{weinberg2019}, and are the consequence of negative radial gas-phase metallicity gradients in the simulated disc during the last $\approx 7$ Gyr, which is when the thin-disc stellar population originated.

Our main conclusions can be summarised as follows.

\begin{enumerate}

\item Our simulated galaxy stellar disc grows from the inside out as a function of time, with a radial gradient in the SFR at any epoch of the galaxy evolution; these results are consistent with the findings of previous studies using chemical-evolution models and simulations (see, for example, \citealt{vincenzo2018b}; \citealt{molla2019} and references therein; \citealt{tissera2019}). The inside-out has the following two effects: first of all, inner regions always have systematically higher SFRs than the outer regions, mostly because of an exponential surface gas density profile which is preserved, scaling with the half-mass radius of galaxy (see Fig. \ref{fig2}a); this is the first important ingredient that can drive and maintain negative radial metallicity gradients on the simulated galaxy disc \citep{molla2019}. Secondly, the onset of star formation in the outer galaxy disc ($d/d_{1/2} > 1.5$) is delayed at much later times than the inner stellar disc (see Fig. \ref{fig2}b). 

    \item We find that the present-day gaseous disc is composed of two main components: \textit{(i)} a radial inflow component of relatively metal-rich and high-[$\alpha$/Fe] star-forming gas, more important for the chemical evolution of the inner galaxy regions, and \textit{(ii)} an accreted component of metal-poor and relatively low-[$\alpha$/Fe] gas, more important for the chemical evolution of the outer regions, which steepens the radial metallicity gradient in the last $\approx 2\;\text{Gyr}$ (see Fig. \ref{fig:gas-flow_thin-disc}). The accretion causes a temporal decrease of [Fe/H] and increase of [$\alpha$/Fe]. Finally, the radial inflow component has an average velocity of $\approx 0.7$ km/s directed inwards; our findings are in agreement with the assumptions of previous works \citep[e.g.,][]{laceyfall1985, portinari2000,bilitewski2012,grisoni2018}. 
    
    \item Although scattered, we predict that the radial profile of the ratio between the outflow and the infall rates of metals does not depend on redshift in our simulated halo (see Fig. \ref{fig:flow_redshift}). In particular, the outer galaxy regions have less efficient metal retention (namely, higher metal outflow-to-infall ratios) than the inner galaxy regions. Since there are always negative gas-phase metallicity gradients on the galaxy disc, this is the second important ingredient which can drive and maintain radial metallicity gradients on the simulated galaxy disc. The time independence of the radial profile stems from the self-similar growth of the DM halo as a function of time, which regulates the rates of outflow and infall of matter as a function of radius.
   
        \item Stellar migrations mostly involve old and metal-rich stars migrating outwards, but there is also a minor component of relatively young and metal-poor stars migrating inwards (see Fig. \ref{fig:migration-age}). The main effect of stellar migration is to flatten the stellar metallicity gradients by 0.05 dex/kpc in the predicted slope (see Fig. \ref{fig:feh-slope}). The typical time scale over which stellar migration takes place is of the order of $\approx 5$ Gyr, being more important for present-day radii corresponding to the solar neighbourhood. Finally, the effect of migrations can appear more in the [$\alpha$/Fe] versus \textit{age} or [$\alpha$/Fe] versus [Fe/H] relations than in [Fe/H] versus \textit{age} (see Fig. \ref{fig:ofe-feh-age-migration2}).

\end{enumerate}

Our simulation can reproduce the observed slope in the radial [$\alpha$/H] and [Fe/H] gradients as observed by APOGEE-DR16, which are also consistent with the Cepheid observations. However, for the gas and in the youngest stellar populations, we predict a steeper slopes at redshift $z=0$ than at $z\approx0.25$ (ages of $\approx3$ Gyr), which is caused by an accretion event of metal-poor gas in the last $2$ Gyr, that triggered star-formation activity at the solar neighourhood and larger radii.
The slope at ages $<2$ Gyr is steeper than those observed in Cepheids and young open clusters \citep{luck2011,korotin2014,genovali2015, magrini2017}, which may constrain the recent gas accretion history.
Finally, despite the success of reproducing the bimodal distribution of thick- and thin-disc stars in [$\alpha$/Fe]--[Fe/H] from APOGEE-DR16, there are interesting mismatches in the dispersion and at the highest and lowest metallicities that deserve a future study, by exploring different sub-grid physics assumptions.

%%%%%%%%%%%%%%%%%%%%%%%%%%%%%%%%%%%%%%%%%%%
%%%%%%%%%%%%%%%%%%%%%%%%%%%%%%%%%%%%%%%%%%%

\section*{Acknowledgments}
We thank the referee, Patricia Beatriz Tissera, for many constructive comments, which improved the quality and clarity of our work. We thank Michael Fall, David Weinberg, Andrea Miglio, and James Johnson for many discussions, Emily Griffith for providing information and useful suggestions about APOGEE-DR16, Anna Porredon, Yi-Kuan Chiang, and Francesco Belfiore for useful remarks, and Volker Springel for providing the code \textsc{Gadget-3}. 
FV acknowledges the support of a Fellowship from the Center for Cosmology and AstroParticle Physics at The Ohio State University. CK acknowledges funding from the United Kingdom Science and Technology Facility Council (STFC) through grant ST/R000905/1. This work used the DiRAC Data Centric system at Durham University, operated by the Institute for Computational Cosmology on behalf of the STFC DiRAC HPC Facility (www.dirac.ac.uk). This equipment was funded by a BIS National E-infrastructure capital grant ST/K00042X/1, STFC capital grant ST/K00087X/1, DiRAC Operations grant ST/K003267/1 and Durham University. DiRAC is part of the National E-Infrastructure.
This research has also made use of the University of Hertfordshire's high-performance computing facility.  

In this work we have made use of SDSS-IV APOGEE-2 DR16 data. Funding for the Sloan Digital Sky Survey IV has been provided by the Alfred P. Sloan Foundation, the U.S. Department of Energy Office of Science, and the Participating Institutions. SDSS-IV acknowledges
support and resources from the Center for High-Performance Computing at
the University of Utah. The SDSS web site is www.sdss.org.

SDSS-IV is managed by the Astrophysical Research Consortium for the 
Participating Institutions of the SDSS Collaboration including the 
Brazilian Participation Group, the Carnegie Institution for Science, 
Carnegie Mellon University, the Chilean Participation Group, the French Participation Group, Harvard-Smithsonian Center for Astrophysics, 
Instituto de Astrof\'isica de Canarias, The Johns Hopkins University, Kavli Institute for the Physics and Mathematics of the Universe (IPMU) / 
University of Tokyo, the Korean Participation Group, Lawrence Berkeley National Laboratory, 
Leibniz Institut f\"ur Astrophysik Potsdam (AIP),  
Max-Planck-Institut f\"ur Astronomie (MPIA Heidelberg), 
Max-Planck-Institut f\"ur Astrophysik (MPA Garching), 
Max-Planck-Institut f\"ur Extraterrestrische Physik (MPE), 
National Astronomical Observatories of China, New Mexico State University, 
New York University, University of Notre Dame, 
Observat\'ario Nacional / MCTI, The Ohio State University, 
Pennsylvania State University, Shanghai Astronomical Observatory, 
United Kingdom Participation Group,
Universidad Nacional Aut\'onoma de M\'exico, University of Arizona, 
University of Colorado Boulder, University of Oxford, University of Portsmouth, 
University of Utah, University of Virginia, University of Washington, University of Wisconsin, 
Vanderbilt University, and Yale University.

%%%%%%%%%%%%%%%%%%%%%%%%%%%%%%%%%%%%%%%%%%%%%%%%%%

\label{lastpage}


\begin{thebibliography}{}

\bibitem[\protect\citeauthoryear{Ahumada, et al.}{2019}]{ahumada2019} Ahumada R., et al., 2019, preprint (arXiv:1912.02905)

\bibitem[Allen et al.(2008)]{allen2008} Allen, M.~G., Groves, B.~A., Dopita, M.~A., Sutherland, R.~S., \& Kewley, L.~J.\ 2008, \apjs, 178, 20

\bibitem[\protect\citeauthoryear{Amarante, et al.}{2020}]{amarante2020} Amarante J.~A.~S., Beraldo e Silva L., Debattista V.~P., Smith M.~C., 2020, ApJL, 891, L30


\bibitem[\protect\citeauthoryear{Anders, et al.}{2017a}]{anders2017a} Anders F., et al., 2017, \aap, 597, A30

\bibitem[\protect\citeauthoryear{Anders, et al.}{2017b}]{anders2017b} Anders F., et al., 2017, \aap, 600, A70

\bibitem[\protect\citeauthoryear{Antoja, et al.}{2018}]{antoja2018} Antoja T., et al., 2018, Natur, 561, 360

\bibitem[\protect\citeauthoryear{Belfiore, et al.}{2019}]{belfiore2019} Belfiore F., Vincenzo F., Maiolino R., Matteucci F., 2019, MNRAS, 487, 456

\bibitem[\protect\citeauthoryear{Belokurov, et al.}{2018}]{belokurov2018} Belokurov V., Erkal D., Evans N.~W., Koposov S.~E., Deason A.~J., 2018, MNRAS, 478, 611

\bibitem[\protect\citeauthoryear{Belokurov, et al.}{2019}]{belokurov2019} Belokurov V., Sanders J.~L., Fattahi A., Smith M.~C., Deason A.~J., Evans N.~W., Grand R.~J.~J., 2019, preprint (arXiv:1909.04679)

\bibitem[\protect\citeauthoryear{Bensby, Feltzing \& Lundstr{\"o}m}{2003}]{bensby2003} Bensby T., Feltzing S., Lundstr{\"o}m I., 2003, \aap, 410, 527

\bibitem[\protect\citeauthoryear{Bensby, Feltzing \& Oey}{2014}]{bensby2014} Bensby T., Feltzing S., Oey M.~S., 2014, \aap, 562, A71

\bibitem[\protect\citeauthoryear{Bignone, Helmi \& Tissera}{2019}]{bignone2019} Bignone L.~A., Helmi A., Tissera P.~B., 2019, ApJL, 883, L5


\bibitem[\protect\citeauthoryear{Bilitewski \& Sch{\"o}nrich}{2012}]{bilitewski2012} Bilitewski T., Sch{\"o}nrich R., 2012, MNRAS, 426, 2266

\bibitem[\protect\citeauthoryear{Bird, et al.}{2013}]{bird2013} Bird J.~C., Kazantzidis S., Weinberg D.~H., Guedes J., Callegari S., Mayer L., Madau P., 2013, ApJ, 773, 43

\bibitem[\protect\citeauthoryear{Boissier \& Prantzos}{2000}]{boissier2000} Boissier S., Prantzos N., 2000, MNRAS, 312, 398

\bibitem[\protect\citeauthoryear{Brook, et al.}{2004}]{brook2004} Brook C.~B., Kawata D., Gibson B.~K., Freeman K.~C., 2004, ApJ, 612, 894

\bibitem[\protect\citeauthoryear{Brook, et al.}{2012}]{brook2012} Brook C.~B., et al., 2012, MNRAS, 426, 690

\bibitem[\protect\citeauthoryear{Brook, et al.}{2020}]{brook2020} Brook C.~B., Kawata D., Gibson B.~K., Gallart C., 2020, preprint (arXiv:2001.02187)


\bibitem[\protect\citeauthoryear{Buck}{2020}]{buck2020} Buck T., 2020, MNRAS, 491, 5435

\bibitem[\protect\citeauthoryear{Buder, et al.}{2018}]{buder2018} Buder S., et al., 2018, MNRAS, 478, 4513

\bibitem[\protect\citeauthoryear{Casagrande, et al.}{2016}]{casagrande2016} Casagrande L., et al., 2016, MNRAS, 455, 987

\bibitem[\protect\citeauthoryear{Calura \& Menci}{2009}]{calura2009} Calura F., Menci N., 2009, MNRAS, 400, 1347

\bibitem[\protect\citeauthoryear{Cescutti, et al.}{2007}]{cescutti2007} Cescutti G., Matteucci F., Fran{\c{c}}ois P., Chiappini C., 2007, \aap, 462, 943

\bibitem[\protect\citeauthoryear{Cescutti \& Kobayashi}{2017}]{cescutti2017} Cescutti G., Kobayashi C., 2017, \aap, 607, A23

\bibitem[Chabrier(2003)]{chabrier2003} Chabrier, G.\ 2003, \pasp, 115, 763

\bibitem[\protect\citeauthoryear{Chaplin, et al.}{2020}]{chaplin2020} Chaplin W.~J., et al., 2020, preprint (arXiv:2001.04653)


\bibitem[\protect\citeauthoryear{Chiappini, Matteucci \& Gratton}{1997}]{chiappini1997} Chiappini C., Matteucci F., Gratton R., 1997, ApJ, 477, 765

\bibitem[\protect\citeauthoryear{Ciuc{\u{a}}, et al.}{2020}]{ciuca2020} Ciuc{\u{a}} I., Kawata D., Miglio A., Davies G.~R., Grand R.~J.~J., 2020, arXiv, arXiv:2003.03316

\bibitem[\protect\citeauthoryear{Clarke, et al.}{2019}]{clarke2019} Clarke A.~J., et al., 2019, MNRAS, 484, 3476

\bibitem[\protect\citeauthoryear{Di Matteo, et al.}{2019}]{dimatteo2019} Di Matteo P., et al., 2019, \aap, 632, A4

\bibitem[\protect\citeauthoryear{Feltzing, Bowers \& Agertz}{2019}]{feltzing2019} Feltzing S., Bowers J.~B., Agertz O., 2019, preprint (arXiv:1907.08011)

\bibitem[\protect\citeauthoryear{Feuillet, et al.}{2018}]{feuillet2018} Feuillet D.~K., et al., 2018, MNRAS, 477, 2326

\bibitem[\protect\citeauthoryear{Feuillet, et al.}{2019}]{feuillet2019} Feuillet D.~K., et al., 2019, MNRAS, 489, 1742

\bibitem[\protect\citeauthoryear{Few, et al.}{2014}]{few2014} Few C.~G., Courty S., Gibson B.~K., Michel-Dansac L., Calura F., 2014, MNRAS, 444, 3845

\bibitem[\protect\citeauthoryear{Font, et al.}{2020}]{font2020} Font A.~S., et al., 2020, preprint (arXiv:2004.01914)


\bibitem[\protect\citeauthoryear{Fragkoudi, et al.}{2019}]{fragkoudi2019} Fragkoudi F., et al., 2019, MNRAS, 488, 3324

\bibitem[\protect\citeauthoryear{Fragkoudi, et al.}{2019}]{fragkoudi2019b} Fragkoudi F., et al., 2019, preprint (arXiv:1911.06826)


\bibitem[\protect\citeauthoryear{Frankel, et al.}{2018}]{frankel2018} Frankel N., Rix H.-W., Ting Y.-S., Ness M., Hogg D.~W., 2018, ApJ, 865, 96

\bibitem[\protect\citeauthoryear{Frankel, et al.}{2019}]{frankel2019} Frankel N., Sanders J., Rix H.-W., Ting Y.-S., Ness M., 2019, ApJ, 884, 99

\bibitem[\protect\citeauthoryear{Gallart, et al.}{2019}]{gallart2019} Gallart C., Bernard E.~J., Brook C.~B., Ruiz-Lara T., Cassisi S., Hill V., Monelli M., 2019, NatAs, 3, 932

\bibitem[\protect\citeauthoryear{Genovali, et al.}{2015}]{genovali2015} Genovali K., et al., 2015, \aap, 580, A17

\bibitem[\protect\citeauthoryear{Grand, et al.}{2018a}]{grand2018a} Grand R.~J.~J., et al., 2018, MNRAS, 474, 3629


\bibitem[\protect\citeauthoryear{Grand, et al.}{2018b}]{grand2018b} Grand R.~J.~J., et al., 2018, MNRAS, 481, 1726

\bibitem[\protect\citeauthoryear{Grand, et al.}{2019}]{grand2019} Grand R.~J.~J., et al., 2019, MNRAS, 490, 4786

\bibitem[\protect\citeauthoryear{Grand, et al.}{2020}]{grand2020} Grand R.~J.~J., et al., 2020, preprint (arXiv:2001.06009)


\bibitem[\protect\citeauthoryear{Gravity Collaboration, et al.}{2019}]{gravity2019} Gravity Collaboration, et al., 2019, \aap, 625, L10

\bibitem[\protect\citeauthoryear{Griffith, Johnson \& Weinberg}{2019}]{griffith2019} Griffith E., Johnson J.~A., Weinberg D.~H., 2019, ApJ, 886, 84

\bibitem[\protect\citeauthoryear{Grisoni, Spitoni \& Matteucci}{2018}]{grisoni2018} Grisoni V., Spitoni E., Matteucci F., 2018, MNRAS, 481, 2570

\bibitem[Haardt \& Madau(haardt1996)]{haardt1996} Haardt, F., \& Madau, P.\ 1996, \apj, 461, 20 

\bibitem[\protect\citeauthoryear{Hachisu, Kato \& Nomoto}{1996}]{hachisu1996} Hachisu I., Kato M., Nomoto K., 1996, ApJL, 470, L97

\bibitem[\protect\citeauthoryear{Hayden, et al.}{2014}]{hayden2014} Hayden M.~R., et al., 2014, AJ, 147, 116

\bibitem[\protect\citeauthoryear{Hayden, et al.}{2018}]{hayden2018} Hayden M.~R., et al., 2018, \aap, 609, A79

\bibitem[\protect\citeauthoryear{Hayden, et al.}{2019}]{hayden2019} Hayden M.~R., et al., 2019, arXiv, arXiv:1901.07565

\bibitem[\protect\citeauthoryear{Haynes \& Kobayashi}{2019}]{haynes2019} Haynes C.~J., Kobayashi C., 2019, MNRAS, 483, 5123

\bibitem[\protect\citeauthoryear{Helmi, et al.}{2018}]{helmi2018} Helmi A., Babusiaux C., Koppelman H.~H., Massari D., Veljanoski J., Brown A.~G.~A., 2018, Natur, 563, 85

\bibitem[\protect\citeauthoryear{Hinshaw, et al.}{2013}]{hinshaw2013} Hinshaw G., et al., 2013, ApJS, 208, 19

\bibitem[\protect\citeauthoryear{Hopkins}{2015}]{hopkins2015} Hopkins P.~F., 2015, MNRAS, 450, 53


\bibitem[\protect\citeauthoryear{Hou, Prantzos \& Boissier}{2000}]{hou2000} Hou J.~L., Prantzos N., Boissier S., 2000, \aap, 362, 921

\bibitem[\protect\citeauthoryear{Iorio \& Belokurov}{2019}]{iorio2019} Iorio G., Belokurov V., 2019, MNRAS, 482, 3868

\bibitem[\protect\citeauthoryear{Johnson \& Weinberg}{2019}]{johnson2019} Johnson J.~W., Weinberg D.~H., 2019, preprint (arXiv:1911.02598)

\bibitem[Katz(1992)]{katz1992} Katz, N.\ 1992, \apj, 391, 502 

\bibitem[Katz et al.(1996)]{katz1996} Katz, N., Weinberg, D.~H., \& Hernquist, L.\ 1996, \apjs, 105, 19 

\bibitem[\protect\citeauthoryear{Kobayashi, et al.}{1998}]{kobayashi1998} Kobayashi C., Tsujimoto T., Nomoto K., Hachisu I., Kato M., 1998, ApJL, 503, L155

\bibitem[Kobayashi(2004)]{kobayashi2004} Kobayashi C., 2004, \mnras, 347, 740 

\bibitem[\protect\citeauthoryear{Kobayashi, et al.}{2006}]{kobayashi2006} Kobayashi C., Umeda H., Nomoto K., Tominaga N., Ohkubo T., 2006, ApJ, 653, 1145

\bibitem[Kobayashi et al.(2007)]{kobayashi2007} Kobayashi C., Springel V., White S.~D.~M., 2007, \mnras, 376, 1465 

\bibitem[Kobayashi \& Nomoto(2009)]{kobayashi2009} Kobayashi, C., \& Nomoto, K.\ 2009, \apj, 707, 1466 

\bibitem[\protect\citeauthoryear{Kobayashi \& Nakasato}{2011}]{kobayashi2011} Kobayashi C., Nakasato N., 2011, ApJ, 729, 16

\bibitem[Kobayashi et al.(2011)]{kobayashi2011b} Kobayashi C., Karakas A.~I., Umeda H., 2011, \mnras, 414, 3231 

\bibitem[\protect\citeauthoryear{Kobayashi}{2016}]{kobayashi2016} Kobayashi C., 2016, Natur, 540, 205

\bibitem[\protect\citeauthoryear{Kobayashi, Leung \& Nomoto}{2019}]{kobayashi2019} Kobayashi C., Leung S.-C., Nomoto K., 2019, preprint (arXiv:1906.09980)

\bibitem[\protect\citeauthoryear{Khoperskov, et al.}{2018}]{khoperskov2018} Khoperskov S., et al., 2018, preprint (arXiv:1811.09205)

\bibitem[\protect\citeauthoryear{Khoperskov, et al.}{2019}]{khoperskov2019} Khoperskov S., et al., 2019, preprint (arXiv:1910.06335)

\bibitem[\protect\citeauthoryear{Kirby, et al.}{2019}]{kirby2019} Kirby E.~N., et al., 2019, ApJ, 881, 45

\bibitem[\protect\citeauthoryear{Kobayashi}{2014}]{kobayashi2014} Kobayashi C., 2014, IAUS, 298, 167

\bibitem[\protect\citeauthoryear{Kobayashi}{2016}]{kobayashi2016iau} Kobayashi C., 2016, IAUS, 317, 57

\bibitem[\protect\citeauthoryear{Kobayashi, Nomoto \& Hachisu}{2015}]{kobayashi2015} Kobayashi C., Nomoto K., Hachisu I., 2015, ApJL, 804, L24

\bibitem[\protect\citeauthoryear{Komatsu, et al.}{2009}]{komatsu2009} Komatsu E., et al., 2009, ApJS, 180, 330

\bibitem[\protect\citeauthoryear{Korotin, et al.}{2014}]{korotin2014} Korotin S.~A., Andrievsky S.~M., Luck R.~E., L{\'e}pine J.~R.~D., Maciel W.~J., Kovtyukh V.~V., 2014, MNRAS, 444, 3301

\bibitem[Kroupa(2008)]{kroupa2008} Kroupa, P., 2008, Pathways Through an Eclectic Universe, 390, 3 

\bibitem[\protect\citeauthoryear{Kubryk, Prantzos \& Athanassoula}{2015a}]{kubryk2015a} Kubryk M., Prantzos N., Athanassoula E., 2015, \aap, 580, A126

\bibitem[\protect\citeauthoryear{Kubryk, Prantzos \& Athanassoula}{2015b}]{kubryk2015b} Kubryk M., Prantzos N., Athanassoula E., 2015, \aap, 580, A127

\bibitem[\protect\citeauthoryear{Lacey \& Fall}{1985}]{laceyfall1985} Lacey C.~G., Fall S.~M., 1985, ApJ, 290, 154

\bibitem[\protect\citeauthoryear{Laporte, Johnston \& Tzanidakis}{2019}]{laporte2019a} Laporte C.~F.~P., Johnston K.~V., Tzanidakis A., 2019, MNRAS, 483, 1427

\bibitem[\protect\citeauthoryear{Laporte, et al.}{2019a}]{laporte2019b} Laporte C.~F.~P., Minchev I., Johnston K.~V., G{\'o}mez F.~A., 2019, MNRAS, 485, 3134

\bibitem[\protect\citeauthoryear{Laporte, et al.}{2019b}]{laporte2019c} Laporte C.~F.~P., Belokurov V., Koposov S.~E., Smith M.~C., Hill V., 2019, MNRAS.tmp, L161

\bibitem[\protect\citeauthoryear{Lebreton, Goupil \& Montalb{\'a}n}{2014}]{lebreton2014} Lebreton Y., Goupil M.~J., Montalb{\'a}n J., 2014, EAS, 65, 99, EAS....65

\bibitem[\protect\citeauthoryear{Leung \& Bovy}{2019a}]{leung2019a} Leung H.~W., Bovy J., 2019, MNRAS, 483, 3255

\bibitem[\protect\citeauthoryear{Leung \& Bovy}{2019b}]{leung2019b} Leung H.~W., Bovy J., 2019, MNRAS, 489, 2079

\bibitem[\protect\citeauthoryear{Lin, et al.}{2019}]{lin2019} Lin J., et al., 2019, MNRAS.tmp, 2724

\bibitem[\protect\citeauthoryear{Luck \& Lambert}{2011}]{luck2011} Luck R.~E., Lambert D.~L., 2011, AJ, 142, 136

\bibitem[\protect\citeauthoryear{Ma, et al.}{2017a}]{ma2017a} Ma X., Hopkins P.~F., Feldmann R., Torrey P., Faucher-Gigu{\`e}re C.-A., Kere{\v{s}} D., 2017, MNRAS, 466, 4780

\bibitem[\protect\citeauthoryear{Ma, et al.}{2017b}]{ma2017b} Ma X., et al., 2017, MNRAS, 467, 2430


\bibitem[\protect\citeauthoryear{Mackereth, et al.}{2018}]{mackereth2018} Mackereth J.~T., Crain R.~A., Schiavon R.~P., Schaye J., Theuns T., Schaller M., 2018, MNRAS, 477, 5072

\bibitem[\protect\citeauthoryear{Mackereth, et al.}{2019}]{mackereth2019} Mackereth J.~T., et al., 2019, MNRAS, 489, 176

\bibitem[\protect\citeauthoryear{Mackereth \& Bovy}{2020}]{mackereth2020} Mackereth J.~T., Bovy J., 2020, MNRAS, 492, 3631

\bibitem[\protect\citeauthoryear{McWilliam, et al.}{2018}]{macwilliam2018} McWilliam A., Piro A.~L., Badenes C., Bravo E., 2018, ApJ, 857, 97

\bibitem[\protect\citeauthoryear{Magrini, et al.}{2009}]{magrini2009} Magrini L., Sestito P., Randich S., Galli D., 2009, \aap, 494, 95

\bibitem[\protect\citeauthoryear{Magrini, et al.}{2016}]{magrini2016} Magrini L., Coccato L., Stanghellini L., Casasola V., Galli D., 2016, \aap, 588, A91

\bibitem[\protect\citeauthoryear{Magrini, et al.}{2017}]{magrini2017} Magrini L., et al., 2017, \aap, 603, A2

\bibitem[\protect\citeauthoryear{Magrini, et al.}{2018}]{magrini2018} Magrini L., et al., 2018, \aap, 618, A102

\bibitem[\protect\citeauthoryear{Maio \& Tescari}{2015}]{maio2015} Maio U., Tescari E., 2015, MNRAS, 453, 3798

\bibitem[\protect\citeauthoryear{Marconi, Matteucci \& Tosi}{1994}]{marconi1994} Marconi G., Matteucci F., Tosi M., 1994, MNRAS, 270, 35

\bibitem[\protect\citeauthoryear{Martinez-Medina, et al.}{2016}]{martinez-medina2016} Martinez-Medina L.~A., Pichardo B., Moreno E., Peimbert A., 2016, MNRAS, 463, 459

\bibitem[\protect\citeauthoryear{Matteucci \& Greggio}{1986}]{matteucci1986} Matteucci F., Greggio L., 1986, \aap, 154, 279

\bibitem[\protect\citeauthoryear{Matteucci \& Francois}{1989}]{matteucci1989} Matteucci F., Francois P., 1989, MNRAS, 239, 885

\bibitem[\protect\citeauthoryear{Matteucci \& Brocato}{1990}]{matteuccibrocato1990} Matteucci F., Brocato E., 1990, ApJ, 365, 539

\bibitem[\protect\citeauthoryear{Matteucci}{2001}]{matteucci2001} Matteucci F., 2001, ASSL..253

\bibitem[\protect\citeauthoryear{Matteucci \& Recchi}{2001}]{matteuccirecchi2001} Matteucci F., Recchi S., 2001, ApJ, 558, 351

\bibitem[\protect\citeauthoryear{Matteucci}{2012}]{matteucci2012} Matteucci F., 2012, ceg..book

\bibitem[\protect\citeauthoryear{Miglio, et al.}{2013}]{miglio2013} Miglio A., et al., 2013, MNRAS, 429, 423

\bibitem[\protect\citeauthoryear{Miglio, et al.}{2017}]{miglio2017} Miglio A., et al., 2017, AN, 338, 644

\bibitem[\protect\citeauthoryear{Minchev, et al.}{2011}]{minchev2011} Minchev I., Famaey B., Combes F., Di Matteo P., Mouhcine M., Wozniak H., 2011, \aap, 527, A147

\bibitem[\protect\citeauthoryear{Minchev, et al.}{2018}]{minchev2018} Minchev I., et al., 2018, MNRAS, 481, 1645

\bibitem[\protect\citeauthoryear{Moll{\'a}, Ferrini \& D{\'\i}az}{1997}]{molla1997} Moll{\'a} M., Ferrini F., D{\'\i}az A.~I., 1997, ApJ, 475, 519

\bibitem[\protect\citeauthoryear{Moll{\'a}, et al.}{2015}]{molla2015} Moll{\'a} M., Cavichia O., Gavil{\'a}n M., Gibson B.~K., 2015, MNRAS, 451, 3693

\bibitem[\protect\citeauthoryear{Moll{\'a}, et al.}{2016}]{molla2016} Moll{\'a} M., D{\'\i}az {\'A}. I., Gibson B.~K., Cavichia O., L{\'o}pez-S{\'a}nchez {\'A}.-R., 2016, MNRAS, 462, 1329

\bibitem[\protect\citeauthoryear{Moll{\'a}, et al.}{2019a}]{molla2019} Moll{\'a} M., et al., 2019, MNRAS, 482, 3071

\bibitem[\protect\citeauthoryear{Moll{\'a}, et al.}{2019b}]{molla2019b} Moll{\'a} M., et al., 2019, MNRAS, 490, 665

\bibitem[Monaghan(1992)]{monaghan1992} Monaghan, J.~J.\ 1992, \araa, 30, 543

\bibitem[\protect\citeauthoryear{Mott, Spitoni \& Matteucci}{2013}]{mott2013} Mott A., Spitoni E., Matteucci F., 2013, MNRAS, 435, 2918

\bibitem[\protect\citeauthoryear{Nomoto, Iwamoto \& Kishimoto}{1997}]{nomoto1997} Nomoto K., Iwamoto K., Kishimoto N., 1997, Sci, 276, 1378

\bibitem[\protect\citeauthoryear{Pagel}{2009}]{pagel2009} Pagel B.~E.~J., 2009, nceg.book

\bibitem[\protect\citeauthoryear{Pezzulli \& Fraternali}{2016}]{pezzulli2016} Pezzulli G., Fraternali F., 2016, MNRAS, 455, 2308

\bibitem[\protect\citeauthoryear{Pinsonneault, et al.}{2018}]{pinsonneault2018} Pinsonneault M.~H., et al., 2018, ApJS, 239, 32


\bibitem[\protect\citeauthoryear{Portinari \& Chiosi}{2000}]{portinari2000} Portinari L., Chiosi C., 2000, \aap, 355, 929

\bibitem[\protect\citeauthoryear{Prantzos \& Boissier}{2000}]{prantzos2000} Prantzos N., Boissier S., 2000, MNRAS, 313, 338

\bibitem[\protect\citeauthoryear{Prantzos, et al.}{2018}]{prantzos2018} Prantzos N., Abia C., Limongi M., Chieffi A., Cristallo S., 2018, MNRAS, 476, 3432

\bibitem[\protect\citeauthoryear{Recchi, Matteucci \& D'Ercole}{2001}]{recchi2001} Recchi S., Matteucci F., D'Ercole A., 2001, MNRAS, 322, 800

\bibitem[\protect\citeauthoryear{Reddy, et al.}{2003}]{reddy2003} Reddy B.~E., Tomkin J., Lambert D.~L., Allende Prieto C., 2003, MNRAS, 340, 304

\bibitem[\protect\citeauthoryear{Rendle, et al.}{2019}]{rendle2019} Rendle B.~M., et al., 2019, MNRAS, 490, 4465

\bibitem[\protect\citeauthoryear{Renzini \& Andreon}{2014}]{RenziniAndreon14} Renzini A., Andreon S., 2014, MNRAS, 444, 3581

\bibitem[\protect\citeauthoryear{Romano, et al.}{2005}]{romano2005} Romano D., Chiappini C., Matteucci F., Tosi M., 2005, \aap, 430, 491

\bibitem[\protect\citeauthoryear{Romano, et al.}{2010}]{romano2010} Romano D., Karakas A.~I., Tosi M., Matteucci F., 2010, \aap, 522, A32

\bibitem[\protect\citeauthoryear{Sanders \& Das}{2018}]{sanders2018} Sanders J.~L., Das P., 2018, MNRAS, 481, 4093

\bibitem[Scannapieco et al.(2012)]{scannapieco2012} Scannapieco, C., Wadepuhl, M., Parry, O.~H., et al.\ 2012, \mnras, 423, 1726 

\bibitem[\protect\citeauthoryear{Schaefer, et al.}{2020}]{schaefer2020} Schaefer A.~L., Tremonti C., Belfiore F., Pace Z., Bershady M.~A., Andrews B.~H., Drory N., 2020, ApJL, 890, L3


\bibitem[\protect\citeauthoryear{Sch{\"o}nrich \& Binney}{2009}]{schoenrich2009} Sch{\"o}nrich R., Binney J., 2009, MNRAS, 396, 203

\bibitem[\protect\citeauthoryear{Sch{\"o}nrich \& McMillan}{2017}]{schoenrich2017} Sch{\"o}nrich R., McMillan P.~J., 2017, MNRAS, 467, 1154

\bibitem[\protect\citeauthoryear{Sch{\"o}nrich, McMillan \& Eyer}{2019}]{schoenrich2019} Sch{\"o}nrich R., McMillan P., Eyer L., 2019, MNRAS, 487, 3568

\bibitem[\protect\citeauthoryear{Sellwood \& Binney}{2002}]{selwood2002} Sellwood J.~A., Binney J.~J., 2002, MNRAS, 336, 785

\bibitem[\protect\citeauthoryear{Silva Aguirre, et al.}{2018}]{silvaaguirre2018} Silva Aguirre V., et al., 2018, MNRAS, 475, 5487

\bibitem[\protect\citeauthoryear{Spitoni, et al.}{2009}]{spitoni2009} Spitoni E., Matteucci F., Recchi S., Cescutti G., Pipino A., 2009, \aap, 504, 87

\bibitem[\protect\citeauthoryear{Spitoni \& Matteucci}{2011}]{spitoni2011} Spitoni E., Matteucci F., 2011, \aap, 531, A72

\bibitem[\protect\citeauthoryear{Spitoni, et al.}{2015}]{spitoni2015} Spitoni E., Romano D., Matteucci F., Ciotti L., 2015, ApJ, 802, 129

\bibitem[\protect\citeauthoryear{Spitoni, et al.}{2019}]{spitoni2019} Spitoni E., Silva Aguirre V., Matteucci F., Calura F., Grisoni V., 2019, \aap, 623, A60

\bibitem[Springel(2005)]{springel2005} Springel V., 2005, \mnras, 364, 1105

\bibitem[\protect\citeauthoryear{Springel, et al.}{2008}]{springel2008} Springel V., et al., 2008, MNRAS, 391, 1685

\bibitem[\protect\citeauthoryear{Stanghellini \& Haywood}{2010}]{stanghellini2010} Stanghellini L., Haywood M., 2010, ApJ, 714, 1096

\bibitem[\protect\citeauthoryear{Stanghellini \& Haywood}{2018}]{stanghellini2018} Stanghellini L., Haywood M., 2018, ApJ, 862, 45

\bibitem[Sutherland \& Dopita(1993)]{sutherland1993} Sutherland, R.~S., \& Dopita, M.~A.\ 1993, \apjs, 88, 253 


\bibitem[Taylor \& Kobayashi(2014)]{taylor2014} Taylor, P., \& Kobayashi, C.\ 2014, \mnras, 442, 2751 

\bibitem[\protect\citeauthoryear{Tinsley}{1980}]{tinsley1980} Tinsley B.~M., 1980, FCPh, 5, 287

\bibitem[\protect\citeauthoryear{Tissera, et al.}{2019}]{tissera2019} Tissera P.~B., et al., 2019, MNRAS, 482, 2208

\bibitem[\protect\citeauthoryear{Torrey, et al.}{2019}]{torrey2019} Torrey P., et al., 2019, MNRAS, 484, 5587


\bibitem[\protect\citeauthoryear{Valentini, et al.}{2019}]{valentini2019} Valentini M., Borgani S., Bressan A., Murante G., Tornatore L., Monaco P., 2019, MNRAS, 485, 1384


\bibitem[Vincenzo \& Kobayashi(2018a)]{vincenzo2018a} Vincenzo, F., \& Kobayashi, C.\ 2018, \aap, 610, L16

\bibitem[Vincenzo \& Kobayashi(2018b)]{vincenzo2018b} Vincenzo, F., \& Kobayashi, C.\ 2018, \mnras, 478, 155 

\bibitem[\protect\citeauthoryear{Vincenzo, Kobayashi \& Taylor}{2018}]{vkt2018} Vincenzo F., Kobayashi C., Taylor P., 2018, MNRAS, 480, L38

\bibitem[\protect\citeauthoryear{Vincenzo, et al.}{2019a}]{vincenzo2019a} Vincenzo F., Spitoni E., Calura F., Matteucci F., Silva Aguirre V., Miglio A., Cescutti G., 2019, MNRAS, 487, L47

\bibitem[\protect\citeauthoryear{Vincenzo, et al.}{2019b}]{vincenzo2019c} Vincenzo F., Miglio A., Kobayashi C., Mackereth J.~T., Montalban J., 2019, \aap, 630, A125

\bibitem[\protect\citeauthoryear{Vincenzo, Kobayashi \& Yuan}{2019}]{vincenzo2019} Vincenzo F., Kobayashi C., Yuan T., 2019, MNRAS, 488, 4674

\bibitem[\protect\citeauthoryear{Weinberg, et al.}{2019}]{weinberg2019} Weinberg D.~H., et al., 2019, ApJ, 874, 102

\bibitem[\protect\citeauthoryear{Xiang, et al.}{2019}]{xiang2019} Xiang M., et al., 2019, ApJS, 245, 34

\end{thebibliography}
\end{document}